# Negative Magnetization and Magnetic Ordering of Rare Earth and Transition Metal Sublattices in $NdFe_{0.5}Cr_{0.5}O_3$

S. Kanthal[a], A. Banerjee[a,b], S. Chatterjee[c], P. Yanda[d,], A. Sundaresan[d], D. D. Khalyavin[e], F. Orlandi[e],T. Saha-Dasgupta[f#] and S. Bandyopadhyay[a,b*]

[a] Department of Physics, University of Calcutta, 92 A.P.C. Road, Kolkata: 700009, India
[b] CRNN, University of Calcutta, Sector III, Salt Lake, Kolkata: 700106, India
[c] UGC DAE Consortium for Scientific Research, Kolkata Centre, LB-8, Salt Lake, Kolkata: 700098, India
[d] School of Advanced Materials and Chemistry and Physics of Materials Unit, Jawaharlal Nehru Centre Advanced Scientific Research, Bengaluru: 560064, India
[e] ISIS Pulsed Neutron and Muon Source, STFC Rutherford Appleton Laboratory, Harwell Campus, Didcot, Oxon, OX11 0QX, United Kingdom
[f] S N Bose National Centre for Basic Sciences, JD Block, Salt Lake, Kolkata: 700106, India
#E-mail: tanusri@bose.res.in *E-mail: sbaphy@caluniv.ac.in

## Abstract

We investigate the effect of alloying at the 3*d* transition metal site of a rare-earth-transition metal oxide, by considering $NdFe_{0.5}Cr_{0.5}O_3$ alloy with two equal and random distribution of 3*d* ions, Cr and Fe, interacting with an early 4*f* rare earth ion, Nd. Employing temperature- and field-dependent magnetization measurements, temperature-dependent x-ray diffraction, neutron powder diffraction, and Raman spectroscopy, we characterize its structural and magnetic properties. Our study reveals bipolar magnetic switching (arising from negative magnetization) and magnetocaloric effect which underline the potential of the studied alloy in device application. The neutron diffraction study shows the absence of spin reorientation transition over the entire temperature range of 1.5-320 K, although both parent compounds exhibit spin orientation transition. We discuss the microscopic origin of this curious behavior. The neutron diffraction results also reveal the ordering of Nd spins at an unusually high temperature of about 40 K, which is corroborated by Raman measurements.

## I. Introduction

The study of $RTMO_3$ ($ABO_3$) ($R$-4$f$ rare earth, $TM$-3$d$ transition metal) perovskites are in focus for several decades owing to their intriguing properties, the complex magnetism that requires complex spin Hamiltonian description along with the promise towards applications. The coupling between two magnetic sublattices, the rare earth ion with unpaired 4$f$ electrons and transition metal ion with unpaired 3$d$ electrons, is considered to be one of main sources behind several striking phenomenon that are exhibited by these perovskites.

In particular, transition metals Cr and Fe based $RTMO_3$ families have drawn attention for their exotic magnetic phase transitions, that involve spin reorientation and ordering of $TM$ and $R$ ions. The spins at $TM$ sites of orthochromites ($RCrO_3$) and orthoferrites ($RFeO_3$) order antiferromagnetically, with canting resulting in a small ferromagnetic component, below a Neél temperature ($T_N$) and some of them show a spin reorientation transition (SRT) at a temperature ($T_{SRT}$) below $T_N$. The $R$ spins may or may not show ordering at a further lower temperature. The antiferromagnetic (AFM) transition of $TM$ spins, driven by $TM$-O-$TM$ super-exchange interaction, generally happens in the temperature range of 620-740 K for $RFeO_3$ and 120-300K for $RCrO_3$ [1-4]. Small ferromagnetic (FM) component due to canted spin alignment arises from Dzyaloshinskii-Moriya (DM) interactions [5]. In general, $RTMO_3$ type perovskite which falls in *Pnma* space group has eight irreducible representations (specific spin configuration). Out of these eight representations four irreducible representations allow ordering of $TM$ sublattice whereas all eight representations are compatible with the ordering of $R$ sublattice. Following the Bertuat notation, first introduced in 1974, out of those four irreducible representations three spin states- $\Gamma_1(G_x, C_y, A_z)$, $\Gamma_2(C_x, G_y, F_z)$ and $\Gamma_4(A_x, F_y, G_z)$ are observed for the $TM$ metal site in $RTMO_3$. Presence of magnetic rare earth ions makes this system even more interesting as below $T_N$ overall magnetization is governed by the interactions between small ferromagnetic (FM) component of *TM* sublattice and the paramagnetic (PM) moments of *R* ions through 3*d*-4*f* coupling.

Magnetic features like magnetization reversal (MR), SRT, magnetoelectric effect, negative thermal expansion etc., observed in orthochromites and orthoferrites, are the manifestation of these interactions. As the sublattice magnetization of rare earth spins orients antiparallel to the transition metal sublattice magnetization, a compensation point ($T_{comp}$) is achieved and negative magnetization is obtained below $T_{comp}$ [7-9]. One of the manifestations of this negative magnetization is bipolar magnetic switching (BMS), which opens up the potential for the use of such orthoferrites and orthochromites in spintronics applications. For applications in magnetic memory-based storage in spintronic devices and magnetic switches, it is highly desirable to switch between two distinct states of any physical property e.g., magnetization, electrical resistance, reflectivity etc. by application of an external actuator like electric or magnetic field, temperature, light, etc. Unlike FM materials, which are generally used in spintronics applications, where magnetization can be switched between positive and negative values by flipping the polarity of the applied field, in AFM systems, the switching can be achieved via bipolar magnetic switching below compensation temperature by changing the magnetic field [6, 10, 11]. This makes $RTMO_3$ compounds with *R*-*TM* two magnetic sublattices, promising candidates for applications in spintronics via thermomagnetic switches and random-access memory in addition to other multifunctional devices.

Comparison of magnetic properties of the orthochromite and orthoferrite family reveals few interesting contrasts, a) availability of different magnetic phase ($\Gamma_2$ or $\Gamma_4$) below $T_N$, with either discrete or continuous SRT, are far more versatile in orthochromites than in orthoferrites. b) a drastic difference between $T_N$ values for orthoferrites and orthochromites, and nature of SRT, hinting for largely different *d-d* and *f-d* interactions. In view of this contrast, one would expect the properties like bipolar magnetic switching, magnetocaloric effect, and SRT can be tuned by alloying orthochromites and orthoferrites. Indeed, few studies on such mixed compounds, explored MR and reported temperature dependent magnetic spin textures. Some examples are $LaFe_{0.5}Cr_{0.5}O_3$, $YFe_{0.5}Cr_{0.5}O_3$, $NdCr_{1-x}Fe_xO_3$ (x=0.05-0.2) [12-15]. In this study, we focus on $NdFe_{0.5}Cr_{0.5}O_3$ mixed ortho-chromite-ferrite compound. We discuss the rational for our choice in the following.

A thorough literature survey was performed which suggested that all investigated $RFe_{0.5}Cr_{0.5}O_3$ mixed compounds can be classified into five broad categories (Group A-E) based on diversity in spin configuration and spin transitions, as shown in Figure 1.

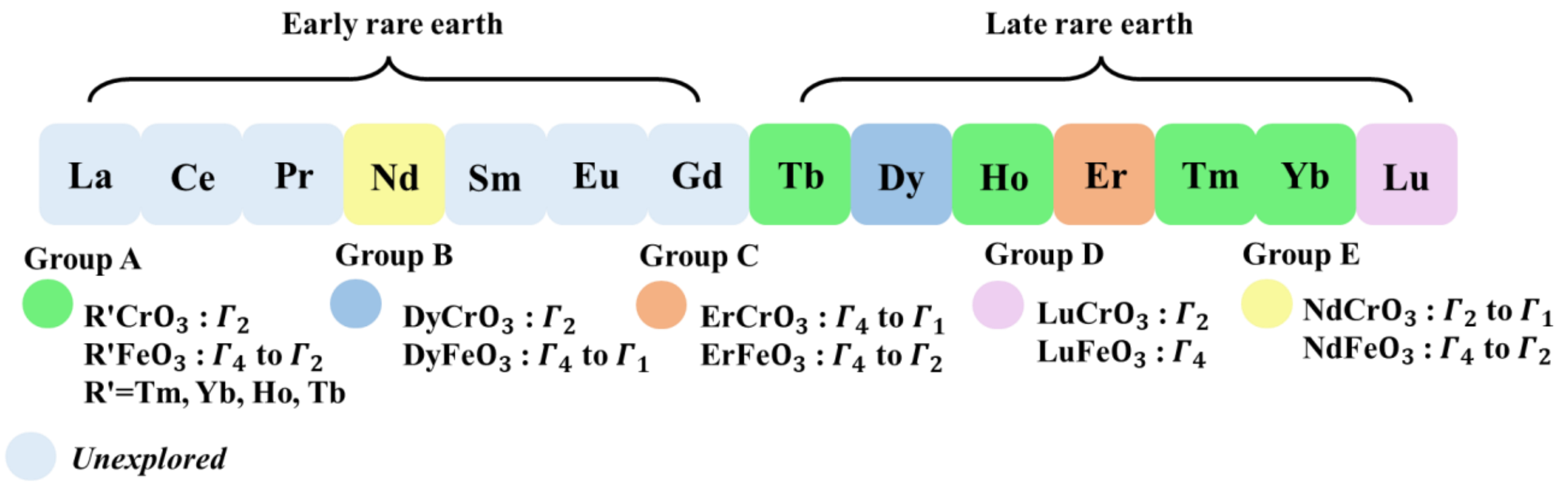

Figure 1: Classification of $RFe_{0.5}Cr_{0.5}O_3$ based on spin ordering, SRT in Fe/Cr and rare earth sublattices. Colour defines a specific group where the spin configurations $\Gamma_1$, $\Gamma_2$ and $\Gamma_4$ ($\Gamma_1$, $\Gamma_2$ and $\Gamma_4$ are irreducible representations according to Bertuat notation [6]) in $RFeO_3$ and $RCrO_3$ are mentioned under that specific group except the unexplored rare earths.

In case of Group A and B there is no SRT associated with $RCrO_3$ but $RFeO_3$ undergoes an SRT. Further, for Group A compounds (*R*= Tm, Yb, Ho, and Tb), the low temperature (below SRT) phase of $RFeO_3$ is same as the phase of $RCrO_3$, while it is not the case for Group B (*R*= Dy).

For Group C ($R$=Er), both $R$CrO$_3$ and $R$FeO$_3$ undergo SRT and high temperature phase for both are same i.e., $\Gamma_4$. All $R$Fe$_{0.5}$Cr$_{0.5}$O$_3$ mixed compounds that fall under any of these three Groups A/B/C undergo continuous SRT where a mixed phase of $\Gamma_4 + \Gamma_2$ exists. So far Lu is the only element which falls in Group D where none of the parents shows SRT. LuCrO$_3$ and LuFeO$_3$ possess $\Gamma_2$ and $\Gamma_4$ phases respectively. The magnetic behavior of the mixed phase compound, LuFe$_{0.5}$Cr$_{0.5}$O$_3$ is dominated by Fe sublattice and the compound shows only $\Gamma_4$ phase below 300 K. It is to be noted that all the above discussed compounds, contain late $R$-ion, with more than half-filled 4$f$ shell. On the contrary, the study on mixed orthochromite-ferrites with early rare earth ion are limited. One such example is NdFe$_{0.5}$Cr$_{0.5}$O$_3$. This compound falls under a unique class which is Group E where both parents show SRT but unlike Group C, where magnetic phase of both parents above SRT are same, in this case high temperature phase above SRT of $R$CrO$_3$ is same as the low temperature phase of $R$FeO$_3$. Further, both of the parent compounds $R$CrO$_3$ and $R$FeO$_3$ show SRT which involves two of these $\Gamma_1$, $\Gamma_2$, and $\Gamma_4$ magnetic phases.

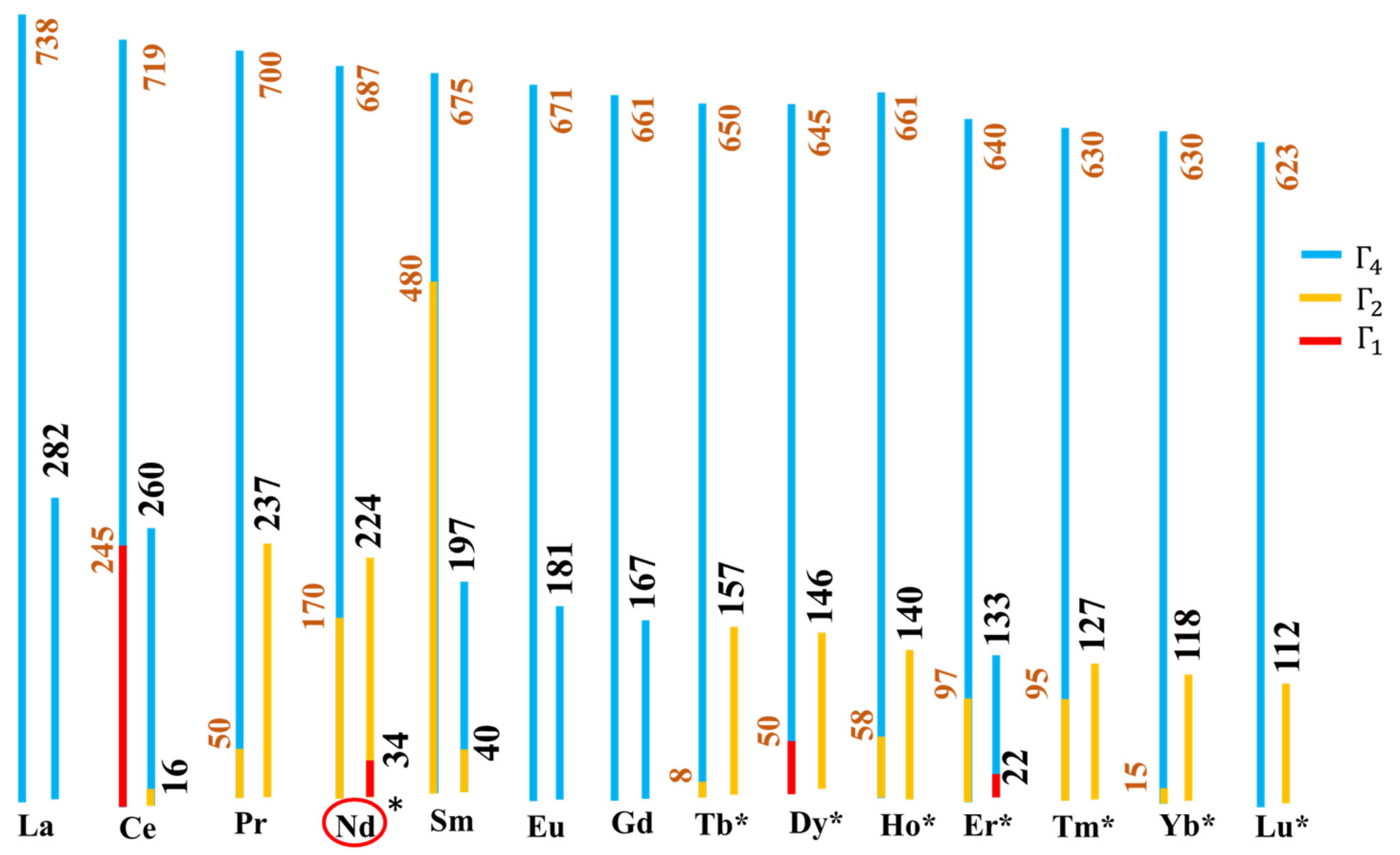

Figure 2: Variation of $T_N$ and $T_{SRT}$ with rare earth metal for $R$FeO$_3$(left) and $R$CrO$_3$ (right) for different choice of $R$-ion. All the temperatures are in Kelvin scale.

The Neel temperature ($T_N$) and $T_{SRT}$ for all $R$FeO$_3$ and $R$CrO$_3$ compounds are shown in Figure 2 with $R$FeO$_3$ ($R$CrO$_3$) shown in left (right) for each rare earth. The $R$ elements, marked with asterisks in Figure 2, indicate the $R$ elements for which the magnetic properties of mixed compounds, $R$Fe$_{0.5}$Cr$_{0.5}$O$_3$ have been studied. As mentioned above, most of the studied cases, include $R$ metals belonging to the late rare-earth series i.e., Tb, Dy, Ho, Er, Tm, Yb, and Lu [16, 17], with the exception of only one study involving Nd-ion [18]. Selection of Nd, appears interesting as parent compounds $NdCrO_3$ and $NdFeO_3$ have distinctly different AFM ordering temperatures of 224 K and 687 K respectively, and below which $NdCrO_3$ acquires $\Gamma_2$ spin configuration, whereas it is $\Gamma_4$ for $NdFeO_3$. Again, there is a SRT from $\Gamma_2$ to $\Gamma_1$ in Cr-sublattice for $NdCrO_3$ at 34 K and it happens continuously from $\Gamma_4$ to $\Gamma_2$ for $NdFeO_3$ in the temperature range 100-170 K. So, below room temperature both $NdCrO_3$ and $NdFeO_3$ have two accessible magnetic states for transition metal sublattice i.e., both individually shows SRT. Thus, one could expect a temperature dependent complex magnetic phase diagram in $NdFe_{0.5}Cr_{0.5}O_3$ similar to that of Dy and Er based compounds. Surprisingly, according to the only available report [18], the mixed compound $NdFe_{0.5}Cr_{0.5}O_3$ exhibit only $\Gamma_2$ phase with no SRT and no ordering of $R$ sublattice was evident. However no scientific explanation was given in that report [18] for the mysterious behavior of $NdFe_{0.5}Cr_{0.5}O_3$. In this study we revisit this issue to find the reason behind this trend using the study of negative magnetization, magnetocaloric effect, x-ray powder diffraction (XRPD), neutron powder diffraction (NPD), Raman spectroscopy, and SQUID magnetometry. Going beyond the findings in previous report [18], our study unravels a) bipolar magnetic switching arising due to the negative magnetization, b) moderate magnetocaloric effect, which together with observation of BMS makes $NdFe_{0.5}Cr_{0.5}O_3$ compound a potential candidate for spintronic device application, c) ordering of Nd spins around 40 K, in contrast to the findings in previous report [18]. The rather high temperature ordering of *R* spins in the mixed compound is interesting. We note that the Nd spins ordering temperature in the mixed compound is much higher than that in the parent phases, which is 10K in $NdCrO_3$ [19, 20] and 1.5 K in $NdFeO_3$ [21-23]. However, this is in line with much higher *R* spin ordering temperature reported for Tm or Ho mixed Cr-Fe compounds, compared to their parent phases [16, 17]. d) existence of strong spin-phonon coupling both below $T_N$ and below the Nd spin ordering temperature. e) the microscopic origin for the mixed compound $NdFe_{0.5}Cr_{0.5}O_3$ exhibiting only $\Gamma_2$ phase with absence of SRT.

## II. Experimental Details

Bulk polycrystalline powder of $NdFe_{0.5}Cr_{0.5}O_3$ was prepared by solid-state reaction method where $Nd_2O_3$, $Cr_2O_3$ and $Fe_2O_3$ were taken in 2:1:1 ratio before grounding them properly in an agate mortar to ensure the homogeneity of the mixture. Then the powder was placed in an alumina crucible and annealed at 1200 ℃ for 24 hr through several intermediate grindings. In the final stage, a pellet formed by pressing the powder was sintered at 1400 ℃ for 24 hr before grounding it into the required polycrystalline powder.

Temperature-dependent x-ray powder diffraction (XRPD) data was collected on a surface detector at Raja Ramanna Centre for Advanced Technology (RRCAT), Indore, India with a wavelength of $\lambda$ = 0.7779 Å. Neutron powder diffraction (NPD) measurements were carried out over a wide temperature range of 1.5 to 320 K at the WISH high resolution cold diffractometer, Rutherford Appleton Laboratory (RAL), UK [24]. Around 2.68 g of fine powder sample was mounted in the standard vanadium can of 6 mm diameter. The vanadium can was loaded inside a liquid He cryostat for low-temperature data acquisition down to 1.5 K. Favorable neutron scattering to absorption cross sections for the constituent magnetic atoms- Nd, Cr, and Fe, helped in high quality data acquisition process. XRPD and NPD patterns were refined by using Rietveld method [25] in FullProf [26] and JANA2006 software package [27] respectively. NPD data was utilized to obtain the temperature evolution of crystallographic and magnetic structure of the sample.

Temperature (*T*) and magnetic field (*H*) dependent DC magnetization data were collected using a Superconducting Quantum Interference Device Vibrating Sample Magnetometer (SQUID VSM Quantum Design MPMS-7 T). Two measuring protocols viz. zero field cool (ZFC) and field cool (FC) were followed during temperature-dependent magnetization (*M*-*T*) measurement and data were recorded during heating cycle. Isothermal field dependent magnetization (*M*-*H*) measurements were performed at selected temperatures up to a field variation of ±5 T. Besides, several virgin isothermal field against magnetization (*M*-*H*) data were collected from 0 to 5 T. Bipolar magnetic switching measurement was also performed at 5 and 10 K.

Temperature dependent Raman spectroscopic data were collected in Renishaw Micro Raman Spectroscopy instrument. The sample plate was mounted and vacuum sealed inside a cryostat system which operates in the temperature range of 10 to 300 K. The sample was illuminated with a 785 nm laser source and the spectrum was recorded from 100 $cm^{-1}$ to 900 $cm^{-1}$ in reflection geometry.

## III. Results and Discussions

### A. Crystal Structure Analysis

The crystal structure of the sample was analyzed using NPD data that were collected from 1.5 to 320 K. Diffraction patterns at 1.5, 50, and 320 K along with Rietveld refined profiles are shown in Figure 3. All crystallographic parameters defining the structure of the compound and reliability parameters of refinement are listed in Table-I. Refinements suggest the orthorhombic crystal structure with a centrosymmetric *Pnma* space group. No impurity peaks were observed, confirming the purity of the sample at least within the sensitivity of the diffraction measurements. The refined profiles shown in Figure 3 contain both nuclear and magnetic contributions. The peaks marked with the star (*) symbol represent magnetic contribution in the system which is discussed in the subsequent section. Low values of weighted profile $R$-factor ($R_{wp}$), expected $R$-factor ($R_p$), and goodness-of-fit ($\chi^2$) indicate that the fitted model is in good agreement with the experimental result. Apart from that, a slight increase in lattice parameters with temperature was noticed. The Rietveld refined X-Ray powder diffraction (XRPD) patterns at some selected temperatures have been presented in the supplemental material (SM)[28] as Figure 1(a)-(c). The value of lattice parameters emerging from XRPD analysis (Table 1 of SM) shows close match with the parameters obtained from NPD analysis. Ideally the discrepancy between lattice parameters at a particular temperature should be zero for XRPD and NPD. However due to sensitivity of the diffractometers a little amount of discrepancy arise. The variation of lattice parameters with temperature as extracted from NPD refinement has been presented in Figure 3(b). Refinement of temperature-dependent XRPD data also suggests similar behavior of lattice parameters as shown in Figure 1(e) of SM [28].

Table-I: Refined crystallographic parameters for $NdFe_{0.5}Cr_{0.5}O_3$ obtained from NPD analysis.

| **Parameter** | **Temperature** | | |
|---|---|---|---|
| | **1.5 K** | **50 K** | **320 K** |
| a (Å) | 5.53342(4) | 5.53350(3) | 5.53562(4) |
| b (Å) | 7.71722(5) | 7.71755(5) | 7.73194(5) |
| c (Å) | 5.42873(4) | 5.42882(2) | 5.43880(4) |
| V (Å$^3$) | 231.821(3) | 231.837(5) | 232.787(3) |
| **Nd: 4c (x, 0.25, z)** | | | |
| x (Å) | 0.04516(2) | 0.04487(4) | 0.04386(3) |
| z (Å) | -0.01077(5) | -0.00974(2) | -0.00918(2) |
| $B_{iso}$ (Å$^2$) | 0.003 | 0.004 | 0.004 |
| **Fe/Cr: 4b (0, 0, 0.25)** | | | |
| $B_{iso}$ (Å$^2$) | 0.005 | 0.006 | 0.006 |
| **O1: 4c (x, 0.25, z)** | | | |
| x (Å) | 0.47731(2) | 0.47711(2) | 0.47785(1) |
| z (Å) | 0.08443(1) | 0.08413(1) | 0.08423(3) |
| $B_{iso}$ (Å$^2$) | 0.007 | 0.008 | 0.008 |
| **O2: 8d (x, y, z)** | | | |
| x (Å) | 0.29334(2) | 0.29320(1) | 0.29250(2) |
| y (Å) | 0.04450(1) | 0.04497(4) | 0.04446(4) |
| z (Å) | -0.29487(2) | -0.29422(2) | -0.29330(3) |
| $B_{iso}$ (Å$^2$) | 0.006 | 0.007 | 0.008 |
| **Reliability parameters** | | | |
| $R_{wp}$ | 5.19 | 5.32 | 4.09 |
| $R_p$ | 5.00 | 5.42 | 3.88 |
| $\chi^2$ | 5.39 | 2.47 | 3.98 |

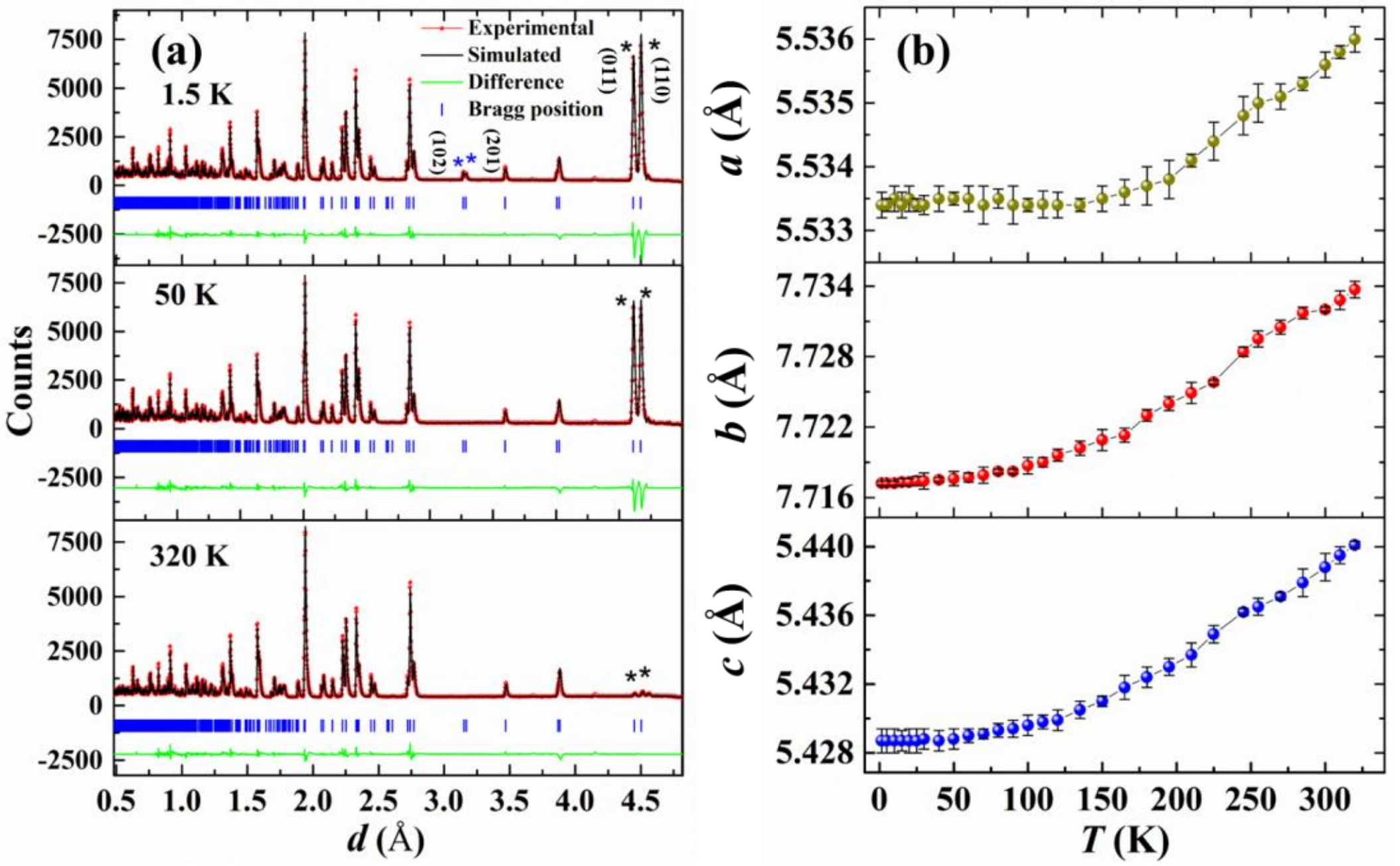

Figure 3: (a) Rietveld refined NPD patterns of $NdFe_{0.5}Cr_{0.5}O_3$ at 1.5, 50 and 320 K respectively. Red circles are the observed intensities with the fitted profile (black line) and the Braggs reflections (vertical lines). (b) The temperature variation of lattice parameters (a, b and c) as obtained from NPD analysis. Error bar is indicated by black vertical lines with each data point.

## B. Magnetic Characterization

**Temperature Dependent Magnetization:**

To characterize the magnetic properties, DC temperature-dependent magnetization (*M-T*) and DC field-dependent magnetization (*M-H*) measurements were performed. Figure 4(a) and (b) show *M-T* data measured under an applied magnetic field of 100 Oe. The magnetic ordering corresponding to Fe/Cr sublattice starts from 244 K (Neél temperature, $T_N$) which supports the previous report [18, 29]. In this mixed *R* based orthochromite and orthoferrite, there exists a random distribution of Cr-O-Fe, Fe-O-Fe, and Cr-O-Cr superexchange paths. Fe-O-Fe as well as Cr-O-Fe superexchange being stronger than Cr-O-Cr, an increase in AFM Neél temperature is expected. Following this expectation, the Neél temperature of $NdFe_{0.5}Cr_{0.5}O_3$ is found to be (~20 K) higher than that of $NdCrO_3$. N. Dasari et. al.,[30] explained the nonlinear behavior of $T_N$ with doping concentration for $YFe_{1-x}Cr_xO_3$ (x = 0.1-0.9) and put forward a mathematical expression which is given by-

$$T_N(x) = \frac{2z}{3k}\left[\sum_{i,j} S_i(S_i+1)S_j(S_j+1)P_i^2P_j^2J_{ij}^2\right]^{\frac{1}{2}} \ldots\ldots(1)$$

where z = nearest-neighbor coordination number, which is 6 for our case. i, j= Cr and Fe, $P_i/P_j$ indicates the probability of site occupancy and $J_{ij}$ is the superexchange interaction strength between Fe and Cr. Considering $S_{Fe} = 5/2$, $S_{Cr} = 3/2$, and $J_{ij} = 24$ K (as obtained for $YFe_{1-x}Cr_xO_3$), the value of $T_N$ (0.5) for $NdFe_{0.5}Cr_{0.5}O_3$ comes out to be 238 K which is slightly lower than the experimental value 244 K. This discrepancy may arise due to the fact that $J_{ij}$ for Nd based mixed perovskite might be higher than that of non-magnetic Y based compound as lattice distortion induced by Nd and Y are different.

For $NdFe_{0.5}Cr_{0.5}O_3$, when the temperature is increased in ZFC mode, magnetization decreases abruptly from positive value and becomes zero at $T_{comp1} = 25.5$ K (Figure 4(a)). Thereafter, with further increase in temperature the magnetization remains negative in the temperature range of 25.5 K < T < 150.5 K and it becomes positive at around 150.5 K. The temperature range of 25.5 K < T < 150.5 K where magnetization remains negative is much larger than that of $NdFeO_3$, one of the parent compounds which also shows negative magnetization in ZFC mode [31]. Negative magnetization in ZFC mode is quite a rare incident and there are few reports available in the literature [31-35]. Although, uncompensated spins [31, 32], trapped field in magnetometer [33] or disorder [34] etc., are considered to be the reasons for magnetization reversal, still there are doubt about the origin of it. Hence, it is very important to justify the negative magnetization in ZFC mode as this may appear due to the trapped field in the magnetometer rather than being an intrinsic property of the sample. We have gone through several measurement protocols to identify the actual reason behind this negative magnetization in ZFC mode. We propose a mechanism that focuses on the temperature dependent ferrimagnetic arrangement of *R* and *TM* spins explaining the behavior in ZFC mode. Actually, *R* and *TM* structure are AFM and the reason of the negative magnetization is due to the relative orientation of the weak FM moment of the two sublattices induced by the coupling of the main AFM order parameter. The ordering of Nd sub-lattice is strong in low-temperature regime and gradually weakens with rise in temperature as evident from NPD analysis discussed in the following section. The locally induced field at Nd site due to Fe/Cr sublattice is responsible for the antiparallel alignment of Nd and Fe/Cr sublattices. Further, *M-T* data suggests that FM vector of Nd sublattice is most likely oriented parallel to the external field. $T_{comp1}$ and $T_{comp2}$ are two compensation temperatures where the overall magnetization becomes zero under ZFC condition as presented in Figure 4(a). In the studied system, the coupling between *TM* and *R* sublattices is antiferromagnetic and the magnetic moment of individual sublattice, as obtained from NPD analysis, largely depends on the temperature. Starting from low-temperature of 5 K which is well below the ordering temperature of Nd (40 K), the ordered moment of Nd-sublattice is fully saturated and dominates over $Fe^{3+}/Cr^{3+}$ moment, making the overall magnetization positive. Actually external field creates some ordered domains of Nd spins which overcomes antiferromagnetic contribution of Fe/Cr sublattice. However, with the increase in temperature the strength of the ordered moments of Nd gradually reduces, resulting in a first compensation temperature ($T_{comp1}$ = 25.5 K) where moments arising from two different sublattices compensate each other. Beyond this point overall magnetization becomes negative, resulting from competition of Nd and $Fe^{3+}/Cr^{3+}$ moments. The $Fe^{3+}/Cr^{3+}$ moment however is too small to re-balance the domain structure, inherited from the low temperature state. Upon further increase in temperature overall magnetization is dominated by $Fe^{3+}/Cr^{3+}$ moment, and another compensation temperature is met ($T_{comp2}$ = 150.5 K) where coercive field becomes nearly 100 Oe and sufficient amount of domains switch to positive value. So, in a wide temperature range from 25.5 to 150.5 K, the magnetization remains negative for the studied compound of $NdFe_{0.5}Cr_{0.5}O_3$, whereas in one of the parent compound $NdFeO_3$ the range of negative magnetization is comparatively narrow (from 7.6 to 29 K) [31]. This whole scenario is explained with the help of the relative spin orientation of two sublattices in Figure 4(a). The sample $NdFe_{0.}5Cr_{0.5}O_3$ shows negative magnetization under FC condition as shown in Figure 4(b). Earlier the phenomenon of negative magnetization was observed in many perovskite oxides [9, 16, 17, 36]. One of the parent compounds of the studied system i.e., $NdFeO_3$ also exhibits magnetization reversal (MR) and spin-switching phenomenon which results from the relative AFM alignment of Fe and Nd sub-lattices [31]. There are other $RFe_{0.5}Cr_{0.5}O_3$ type systems with $R$ = Tm, Lu [16], and Y [9] that also exhibit negative magnetization in FC mode.

In contrast, in FC condition, there is only one compensation point ($T'_{comp1}$) [Figure 4(b) and inset of Figure 4(b)]. In FC condition the system is cooled down from its paramagnetic phase and as it is cooled below the ordering temperature (244 K) the ferromagnetic component of Fe/Cr sublattice gets locked in the direction of the external magnetic field. At the lowest temperature, Nd-moments are fully ordered and Nd-sublattice dominates over Fe/Cr sublattice magnetization resulting an overall negative magnetic moment due to antiferromagnetic coupling between the two sublattices. With rise in temperature the ordering of Nd-sublattice gradually weakens resulting in the decrease in the absolute value of the magnetization. Magnetization becomes positive after crossing compensation point, $T'_{comp1}$where Nd and Fe/Cr sublattice magnetizations compensate each other. This compensation temperature, marked as $T'_{comp1}$ [Figure 4(b)] is largely dependent on the external field in the cooling cycle ($H_{\mathrm{Ext}}$). Inset of Figure 4(b) shows that $T'_{comp1}$ shifts leftward with increase in $H_{\mathrm{Ext}}$ in the cooling cycle up to 1500 Oe and beyond that no compensation point can be observed in the *M-T* curve. Both $T_{comp1}$ and $T'_{comp1}$ originate from similar phenomenon where two different sublattice moments compensate each other under two different conditions, ZFC and FC respectively. Under FC, upon gradual increase of $H_{\mathrm{Ext}}$ in the cooling cycle the locking of ferromagnetic component of Fe/Cr sublattice is increased, hence after a threshold $H_{\mathrm{Ext}}$ ($\geq$ 2000 Oe) there exists no negative magnetization. The detailed analysis along with a fitting of induced field at Nd (as discussed in the temperature dependent magnetization data under ZFC condition) site due Fe/Cr sublattice justifies the negative magnetization, and it has been presented in Figure 2 of SM [28].

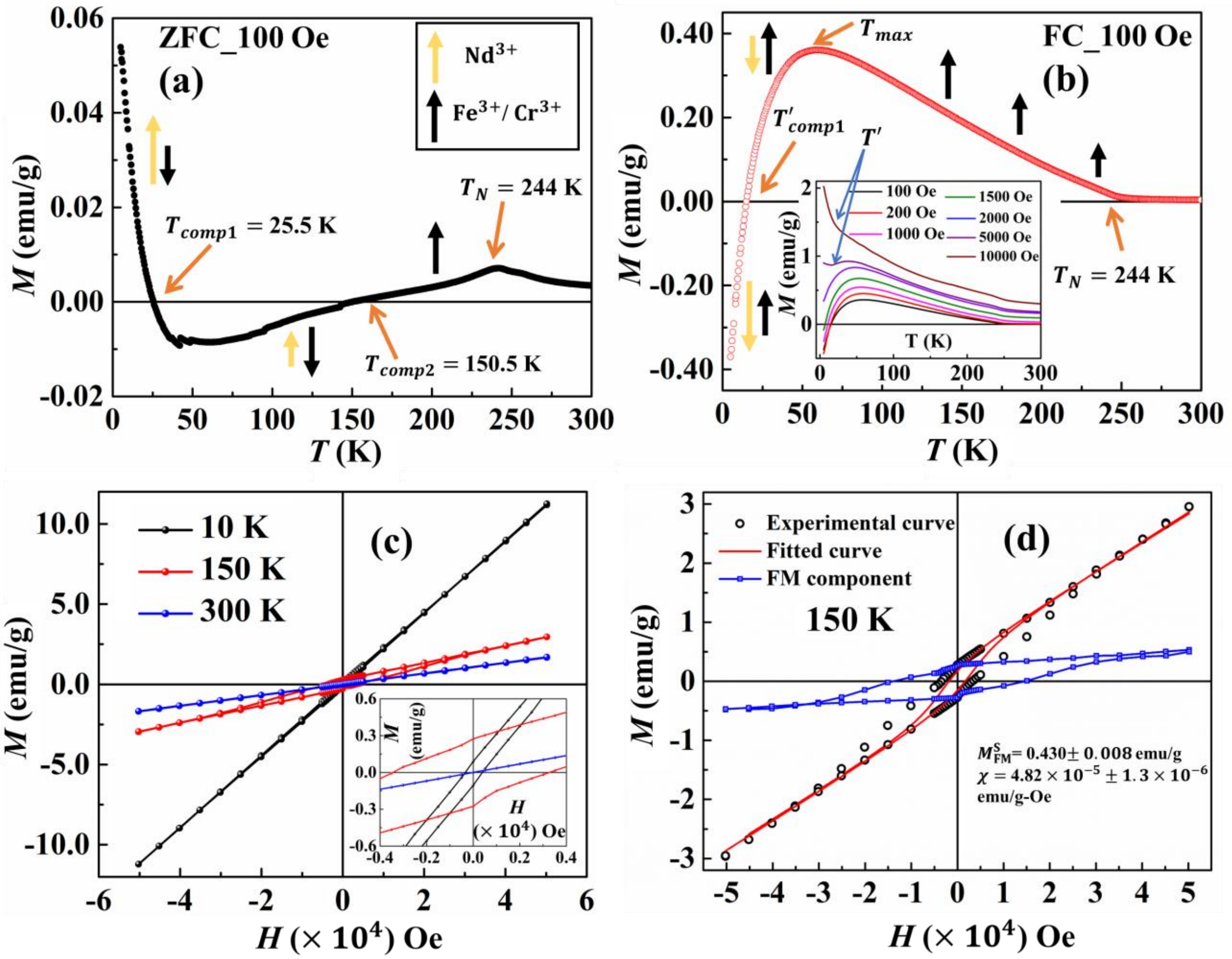

Figure 4: Temperature dependent relative spin orientation under 100 Oe in (a) ZFC protocol (b) FC protocol. Inset of (b) shows field variation of *M*-*T* data. $Fe^{3+}$/$Cr^{3+}$ and $Nd^{3+}$ weak ferromagnetic moments are denoted by black and yellow arrows respectively. (c) *M*-*H* measurement at representative temperatures. (d) Ferromagnetic component of magnetization at 150 K after subtracting paramagnetic contribution. Used symbols in (a)-(d) have been denoted in the text.

**Field Dependent Magnetization:**

*M*-*H* data for $NdFe_{0.5}Cr_{0.5}O_3$ were collected at 10, 150, and 300 K as shown in Figure 4(c). At 150 K, the data features strong hysteresis, together with a linear contribution. The nature confirms long range antiferromagnetic ordering along with the influence of DM interactions in *TM* sublattice, resulting in a canted AFM structure that introduces a small ferromagnetic contribution.

The inset of Figure 4(c) shows significant hysteresis at 150 K. This hysteresis is suppressed as the temperature is increased. *M-H* curve is completely linear at 300 K, indicating paramagnetic nature. Moreover, compared to 150 K, the magnetic hysteresis is relatively weaker at 10 K. The nature of *M-T* curves under FC condition suggests that with the decrease in temperature below 50 K, antiferromagnetic ordering of Nd and *TM* (Fe and Cr) sublattices becomes stronger resulting in an overall weaker magnetic hysteresis at 10 K. The canted AFM ordering of spins causes a weak ferromagnetic component of *TM* sublattice at 150 K and *M-H* curve shows stronger magnetic hysteresis. The FM part at 150 K has been separated from overall magnetization using equation [37]:

$$M(H) = \left(\frac{2M_{FM}^{S}}{\pi}\right)\tan^{-1}[(H \pm H_{Ci})/(H_{Ci}\, tan\{(\pi M_{FM}^{R})/(2M_{FM}^{S})\}] + \chi H \quad \text{.....(2)}$$

*where* $M_{FM}^{S}$, $M_{FM}^{R}$, $H_{Ci}$, *and* $\chi$ *represent saturation magnetization, remanent magnetization, intrinsic coercivity and paramagnetic susceptibility respectively. The estimated FM contribution at 150 K has been presented in Figure 4(d). The value of saturation magnetization and paramagnetic susceptibility at 150 K comes out to be as 0.430 ±0.008 emu/g and* $4.82 \times 10^{-5} \pm 1.3 \times 10^{-6}$ *emu/g-Oe, respectively.*

Magnetic Phase Diagram:

Magnetic phase diagram as shown in Figure 5 has been reconstructed from the temperature dependent magnetization data under FC mode. Considering temperature and external field as tuning parameters, the relative orientation of Fe/Cr and Nd spins split the magnetic phase diagram into several distinct regions (region A to D) as depicted in Figure 5. We describe the field 2000 Oe to be the critical field, $H_C$ which is sufficient enough to keep the net magnetization positive irrespective of temperature. With further rise in the applied field, it becomes sufficient enough to overcome the anisotropic energy and relative AFM alignment between *R* and *TM* ions can be broken, leading to a new spin alignment as indicated in the phase diagram. The inset of Figure 4(b) shows that a higher field can align the *R* moments along *TM* ferromagnetic component giving rise to a sudden increase in the magnetization value at very low temperature regime as evidenced from 5000 Oe curve where a sharp upturn is observed below 15.7 K. In region A, the ordering of *R*-ions is very weak and it hardly opposes the effective magnetization of *TM* sublattice resulting in an increase in average moment with decrease in temperature. With further decrease in the temperature the system enters into B region where AFM ordering of *R* sublattice with respect to the FM component of *TM* sublattice is considerably significant so that magnetization gradually decreases. We denote $T_{max}$ as the temperature where the magnetization attains its maximum value under FC condition. So, $T_{max}$ (red points in Figure 5) separates region A from region B. Thereafter the system may remain in region B ($H_{Ext} > H_C$) or experience negative magnetization in region C ($H_{Ext} < H_C$) which is attributed to the strong AFM ordering of *R*-ions opposing $H_{Ext}$. These states of relative spin alignment can be realized in low or moderate field range as mentioned in Figure 5. When the system moves from region B to C, it experiences a compensation point of magnetization (net zero magnetization) which are shown in the phase diagram as green points ($T'_{comp1}$). Finally, there is a possibility that the system would enter in region D directly from region A without suffering any negative magnetization and it happens in presence of relatively higher $H_{Ext}$. A sufficiently high field could align the *R*-ions in the field direction breaking the AFM ordering caused by the induced local field from *TM* sublattice which results in a sharp increase in magnetization below certain temperatures as mentioned in the inset of Figure 4(b) by black points ($T'$).

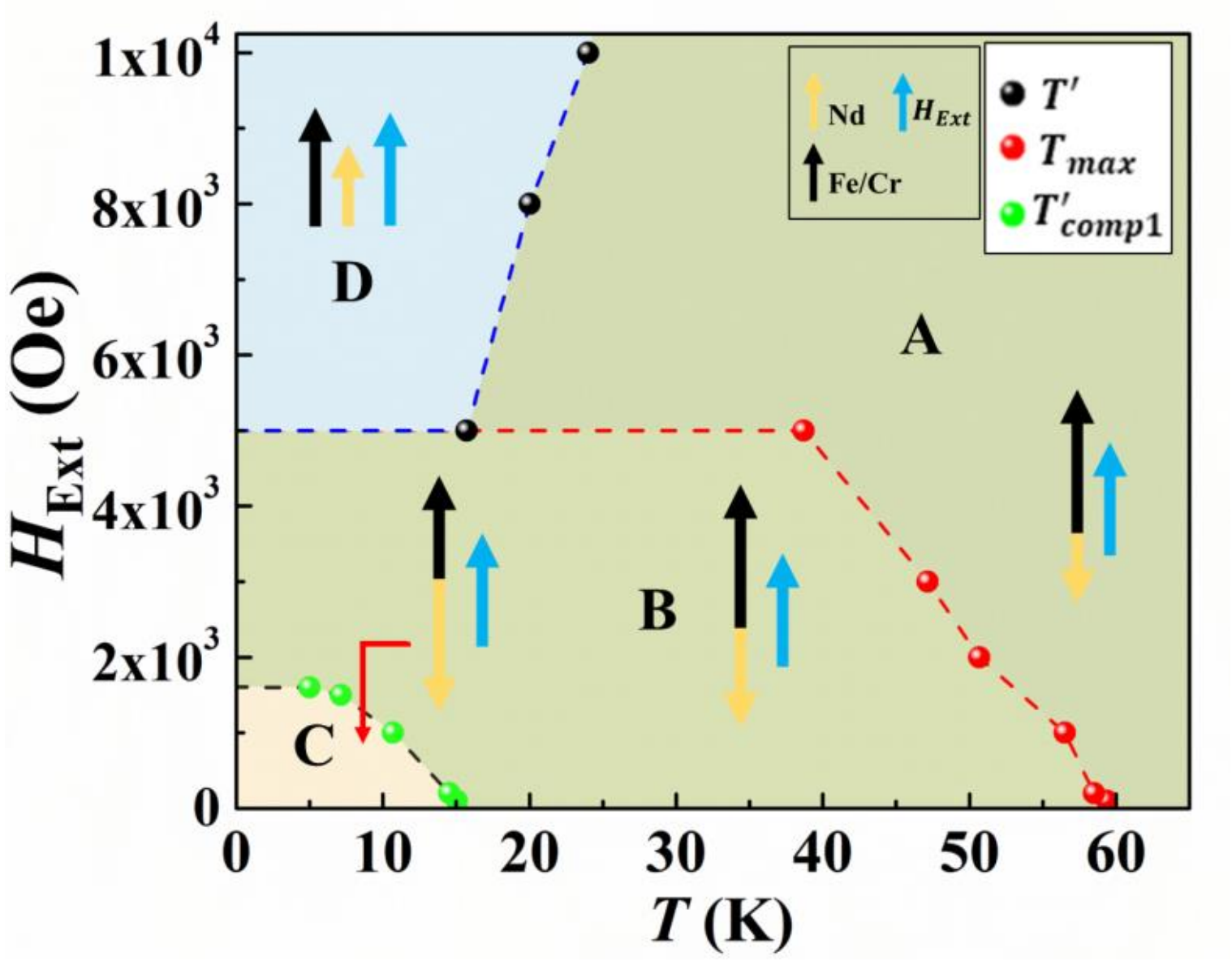

Figure 5: Magnetic phase diagram explaining possible spin evolution of *R* and *TM* sublattices with temperature and field. Black, yellow, and blue arrows define the weak ferromagnetic moment of the $Fe^{3+}$/$Cr^{3+}$spins and, $Nd^{3+}$ spins, and $H_{Ext}$ respectively.

**Bipolar Magnetic Switching:**

During BMS measurement, $NdFe_{0.5}Cr_{0.5}O_3$ was first cooled under 100 Oe field to the desired temperature (< $T'_{comp1}$) and magnetization (-ve) was measured under this field for nearly 8 mins before increasing the field to a higher positive value to obtain an equivalent positive magnetization. This material offers applicability where field-actuated reproducible magnetization switching, as demonstrated in Figure 6, can be utilized. BMS phenomenon was investigated at 5 and 10 K. In both cases, at first, the sample was field cooled under 100 Oe field and then repeating field changing sequences 3000 Oe-100 Oe-3000 Oe-100 Oe… and 1800 Oe-100 Oe-1800 Oe-100 Oe were applied for data acquisition, keeping the temperature fixed at 5 and 10 K respectively. As described in the MR phenomenon, a small $H_{Ext}$ (= $H_1$, here 100 Oe) in FC state produces a stable negative magnetic state with respect to time and a sudden increase of this field to a certain higher value ($H_2$) changes the polarity of magnetization. Thus, a proper selection of temperature, $H_1$ and $H_2$ enable us to utilize this thermomagnetic switching behavior in devices.

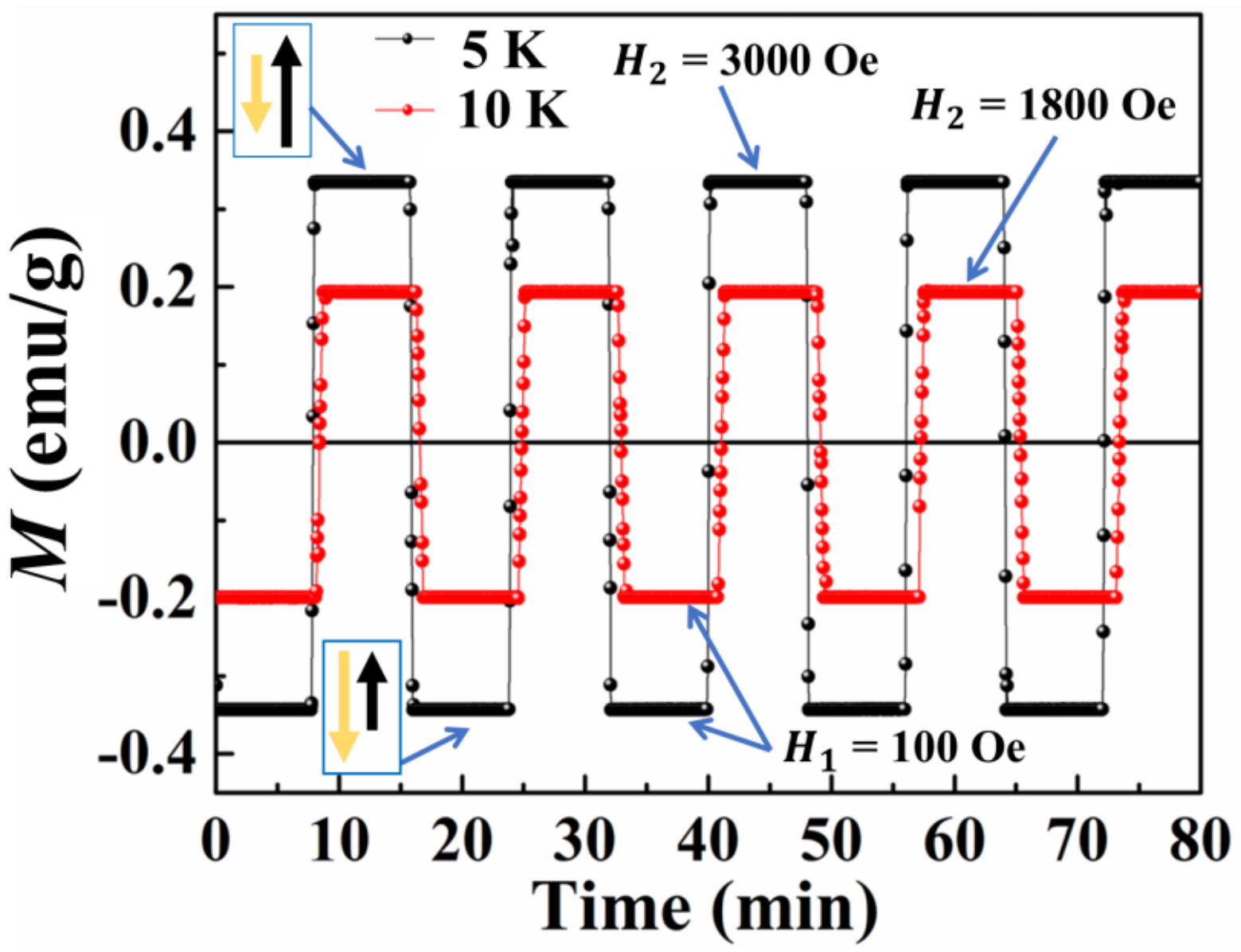

Figure 6: Switching of magnetization at 5 K and 10 K respectively under different field changing sequences based on *M*-*T* curves (Figure 4(b)). The negative magnetized state was obtained by cooling the sample in 100 Oe field. Then application of 3000 Oe and 1800 Oe fields in 5 K and 10 K respectively change the polarity of magnetization. The reciprocity of this switching under field changing sequences was verified for arbitrary time (8 minutes) intervals.

**Magnetocaloric Effect:**

Magnetic entropy change was calculated from isothermal magnetization plots (*M*-*H*), using the Maxwell thermodynamic relation,

$$\Delta S_M = \int_{H_1}^{H_2} \left(\frac{\partial M}{\partial T}\right) dH \ldots.. (3)$$

During the *M*-*H* data acquisition process magnetization was measured for discrete field changes. So, numerical approach was adopted during calculations of the magnetic entropy change following the expression:

$$\Delta S_M(T_i, H_f) = \sum_{J=1}^{f} \frac{M(T_{i+1}, H_j) - M(T_i, H_j)}{T_{i+1} - T_i} \Delta H_j \; \ldots.. \; (4)$$

here $M(T_{i+1})$ and $M(T_i)$ are the initial magnetization measured at $T_{i+1}$ and $T_i$ respectively for a field $H_i$. We have plotted the magnetic entropy change corresponding to an average temperature, $T=(T_{i+1}+T_i)/2$ for different applied fields as depicted in Figure 7.

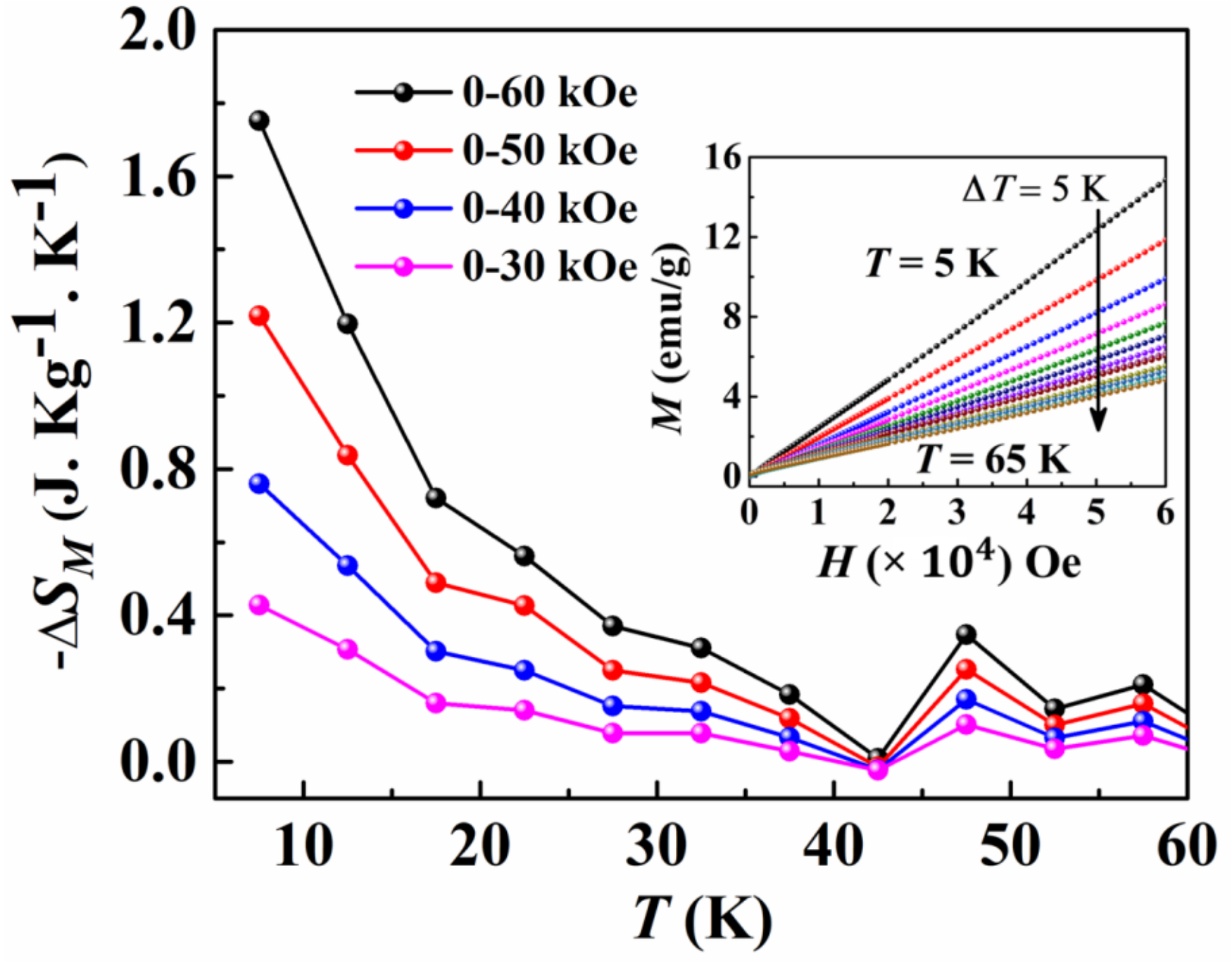

Figure 7:$-\Delta S_M$ versus $T$ plot calculated from isothermal magnetization measurements (Inset shows *M-H* data from 5-65 K with interval of 5 K).

The plot of $-\Delta S_M$ vs $T$ does not show any saturation or peak up to 6 T field. In general, a peak should appear around the temperature where spin reorientation occurs [27, 29]. Absence of such nature indicates the absence of any spin reorientation in that temperature range. From the application point of view, it is desired to have refrigerants that operate in 10 to 80 K temperature range with low magnetic field. Ordering of rare earths at low temperature along with their high magnetic moment help to achieve these requirements. Despite having a relatively low magnetic moment of $Nd^{3+}$ as compared to $Dy^{3+}$ or $Ho^{3+}$, the presence of Cr and Fe in transition metal sublattice in equal proportion enhances the magnetocaloric property as compared to $NdCrO_3$. For our system gradual increase in $-\Delta S_M$ as temperature is decreased from 40 K is due to the ordering of Nd sublattice [16]. The highest value is measured to be $-\Delta S_M$= 1.76 $Jkg^{-1}K^{-1}$ at 6 T and $-\Delta S_M$=1.2 $Jkg^{-1}K^{-1}$ at 5 T which is much higher (~ 371 %) than that of $NdCrO_3$ at 5 T. A previous report where Mn was substituted in $NdCrO_3$ didn't yield much enhancement with respect to magnetocaloric property in $NdCr_{1-x}Mn_xO_3$ (x = 0.1 and 0.15) [27]. So far as the magnetocaloric property of $NdCrO_3$ is concerned this work indicates that strong enhancement can be achieved with Fe incorporation in equal proportion.

**Magnetic Structure of $NdFe_{0.5}Cr_{0.5}O_3$: Absence of SRT and mixed phase:**

The spin configurations of transition metal and rare earth sublattices in $NdFe_{0.5}Cr_{0.5}O_3$ were analyzed using temperature-dependent NPD data. The patterns over the entire temperature range of measurement indicate that the intensity of a few peaks changes significantly with temperature and new peaks appear below a certain temperature. Two Bragg peaks (black * marked in Figure 2(a)) indexed as (110) and (011) signify the long-range magnetic ordering of transition metal sublattice. These two peaks exist for the entire temperature range of measurement with the highest intensities at the lowest temperature (1.5 K). Even at 320 K, the presence of these two peaks (magnetic peaks) can be attributed to the inhomogeneous distribution of Fe-rich clusters. This observation also contributes to the increase in $T_N$ compared to $NdCrO_3$ as observed from *M-T* data. Variations of the intensity of these two peaks are shown in Figure 3(a) of SM [28]. Gradual increase of intensity with a decrease in temperature signifies strengthening of Fe/Cr sublattice ordering. Apart from these two peaks, two new peaks (102) and (201) as shown in Figure 2(a) (blue * marked peaks) appear below 50 K which arises from the ordering of Nd sublattice and this observation was not reported in literature to the best of our knowledge. The temperature evolution of these magnetic peaks is shown in Figure 3(b) of SM.

The analysis of the magnetic structure of $NdFe_{0.5}Cr_{0.5}O_3$ has been performed using the Bertaut notation which labels the magnetic structure with specific ordering along three principal directions [6]. Details of the magnetic space group corresponding to *R* and *TM* sites are described in the SM [28].

Like other $RFe_{0.5}Cr_{0.5}O_3$ [16, 17, 36] type perovskites, magnetic unit cell in $NdFe_{0.5}Cr_{0.5}O_3$ coincides with crystallographic unit cell**,** thus forming a commensurate spin model with propagation vector k = (0, 0, 0). With the resolution of the powder diffraction data, small ferromagnetic component arising from the canted alignment of $Fe^{3+}$/ $Cr^{3+}$ moments cannot be fitted well but *G*-type antiferromagnetic alignment has been analysed properly with an appropriate model. Using ISODISTORT in JANA2006, all mentioned representations in Table-I of SM [28] can be derived and the best agreement with the experimental data was found for $\Gamma_2$ ($Pn'm'a$) i.e., $\Gamma_2(C_x, G_y, F_z)$ representation. Survey of previous reports on $RFe_{0.5}Cr_{0.5}O_3$ suggests that in certain temperature range where the two parent compounds possess different spin configuration, the alloy system generally exhibits the signature of both spin configurations [16, 17, 36]. But in present study for the system of $NdFe_{0.5}Cr_{0.5}O_3$, no such spin configurations were observed. It was found that $\Gamma_2$ return the best fit against the experimental NPD pattern. Figure 8(a)-(c) highlight the fitted NPD pattern for two important magnetic peaks (011) and (110) with $\Gamma_2$ spin configuration at 1.5, 50, and 320 K respectively. Fitted profiles for other temperatures are shown in Figure 4 of SM [28]. The peaks shown in the inset of Figure 8(a) can be attributed to the ordering of the Nd sublattice which starts below 40 K. Analysis revealed $C_x$ type AFM ordering of Nd sublattice. The parent compounds $NdCrO_3$ and $NdFeO_3$ have AFM ordering temperatures of 224 and 687 K respectively and below which $NdCrO_3$ acquires $\Gamma_2$ spin configuration, whereas it is $\Gamma_4$ for $NdFeO_3$. Again, there is a spin reorientation phase transition from $\Gamma_2$ to $\Gamma_1$ for Cr-sublattice in $NdCrO_3$ at 34 K and it happens continuously from $\Gamma_4$ to $\Gamma_2$ for $NdFeO_3$ in the temperature range 100-170 K. Further, Nd sublattice orders in $C_z$ phase below 10 K [19, 20] in $NdCrO_3$ and in $C_y$ phase below 1.5 K in $NdFeO_3$ [21-23]. Hence it seems that the ordering temperature of Nd sublattice is quite high compared to the temperature at which it orders in either of its parent compounds. Similar results have been demonstrated in some previous works related to mixed perovskite with *R* = Tm [16], Ho, Tb, and Er [17]. Particularly, in the case of *R* = Tm and Ho, the *R*-ordering temperature is significantly higher in $R$Fe$_{0.5}$Cr$_{0.5}$O$_3$ than in $R$CrO$_3$ or $R$FeO$_3$. Ho orders at 45 and 35 K in $HoFe_{0.5}Cr_{0.5}O_3$ [17] and $HoFe_{0.4}Cr_{0.6}O_3$ [38] respectively, despite having low ordering temperature in $HoCrO_3$ (7.5 K) [39, 40] and $HoFeO_3$ (4.5 K) [41, 42]. Similarly, in $TmCrO_3$, the ordering temperature of Tm is 6 K [39, 40], whereas for $TmFe_{0.5}Cr_{0.5}O_3$, Tm orders at 52 K [16]. Tb orders at 15 K in $HoFe_{0.5}Cr_{0.5}O_3$, although the ordering temperature of Tb in $TbFeO_3$ was 3 K [17, 43]. However, $RFe_{0.5}Cr_{0.5}O_3$ (R= Dy, Yb) does not show *R* ordering at all [16]. Hence, in some *R* based mixed perovskites ($R$Fe$_x$Cr$_{1-x}$O$_3$) the general trend is that the *R* sublattice in mixed orthochromites or orthoferrites ($R$Fe$_x$Cr$_{1-x}$O$_3$) starts to order at a relatively higher temperature than its normal ordering temperature (in the low temperature regime below or around 10 K) in parent orthochromites ($R$CrO$_3$) or orthoferrites ($R$FeO$_3$). The modulation of *R-R* and *R-TM* interactions strength due to the lattice distortion introduced by two different *TM*-ions, along with the presence of induced field at *R* site due to *TM* sublattice could have a significant impact in *R* ordering.

Figure 8(d)-(f) show the fitting of two main magnetic peaks (011 and 110) with different spin configurations at 245 K where the spin configuration was supposed to follow both parent

compounds. But, $NdFe_{0.5}Cr_{0.5}O_3$ followed neither $\Gamma_4 + \Gamma_2$ nor $\Gamma_4$, but remains in $\Gamma_2$ configuration only.

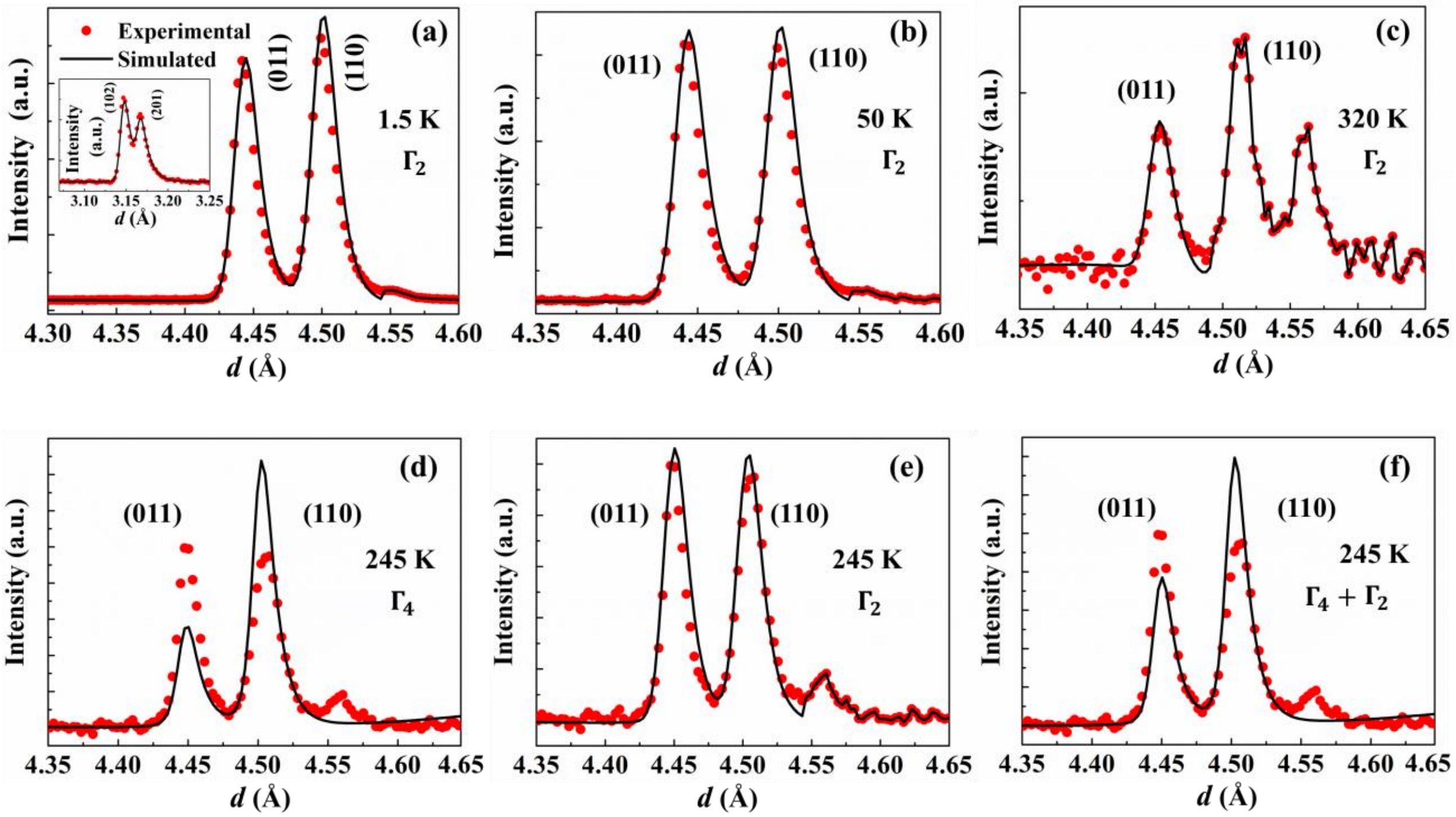

Figure 8: (a)-(c) Magnetic peaks fitted with spin model at three different temperatures 1.5, 50, and 320 K. Inset of (a) presents the magnetic peaks corresponding to Nd sublattice ordering. (d)-(f) Two important magnetic peaks (011) and (110) at 245 K fitted with different spin models.

Therefore, NPD analysis suggests that $NdFe_{0.5}Cr_{0.5}O_3$ does not follow the trend like other so far studied mixed perovskites containing Fe and Cr ($RFe_{0.5}Cr_{0.5}O_3$, $R$ = Tb, Dy, Ho, Er, Yb, and Tm). $NdFe_{0.5}Cr_{0.5}O_3$ does not show any spin reorientation and Fe/Cr sublattice possess only one type of spin configurations i.e., $\Gamma_2$ irrespective of temperature. But below 40 K apart from (011) and (110), two new magnetic Braggs peaks (102) and (201) appear. In absence of any structural transition, these magnetic peaks may appear from the ordering of either $Fe^{3+}$/$Cr^{3+}$ or $Nd^{3+}$ moments. The former possibility can be turned down as both $Fe^{3+}$ and $Cr^{3+}$ occupy same crystallographic site and expected to be ordered at same temperature ($T_N$). So, the only possible reason for the appearance of new peaks (102) and (201) is the ordering of Nd sublattice. Further, we have successfully fitted them considering the ordering of Nd moment in $C_x$ configuration below 40 K. No further change in the magnetic structure is observed upon further cooling down to 1.5 K, apart from the gradual increase in $Nd^{3+}$ moment. Evolution of magnetic moments of transition metal and rare earth sublattices with respect to temperature are shown in Figure 9(a). In order to justify the ordering of Nd sublattice more appropriately, we have plotted integrated intensity of (011)+(110) and (102)+(201) magnetic reflections with temperature in Figure 9(b). Integrated intensity of (011)+(110) magnetic reflections increases below 250 K and almost saturates around 40 K. This signifies the long-range ordering of Fe/Cr sublattice. Below 40 K, we found an increase in integrated intensity of (011)+(110) magnetic reflections as shown in the inset of Figure 9(b). Moreover, temperature dependent integrated intensity of (102)+(201) magnetic reflections also starts to appear below 40 K and that strongly confirms ordering of Nd.

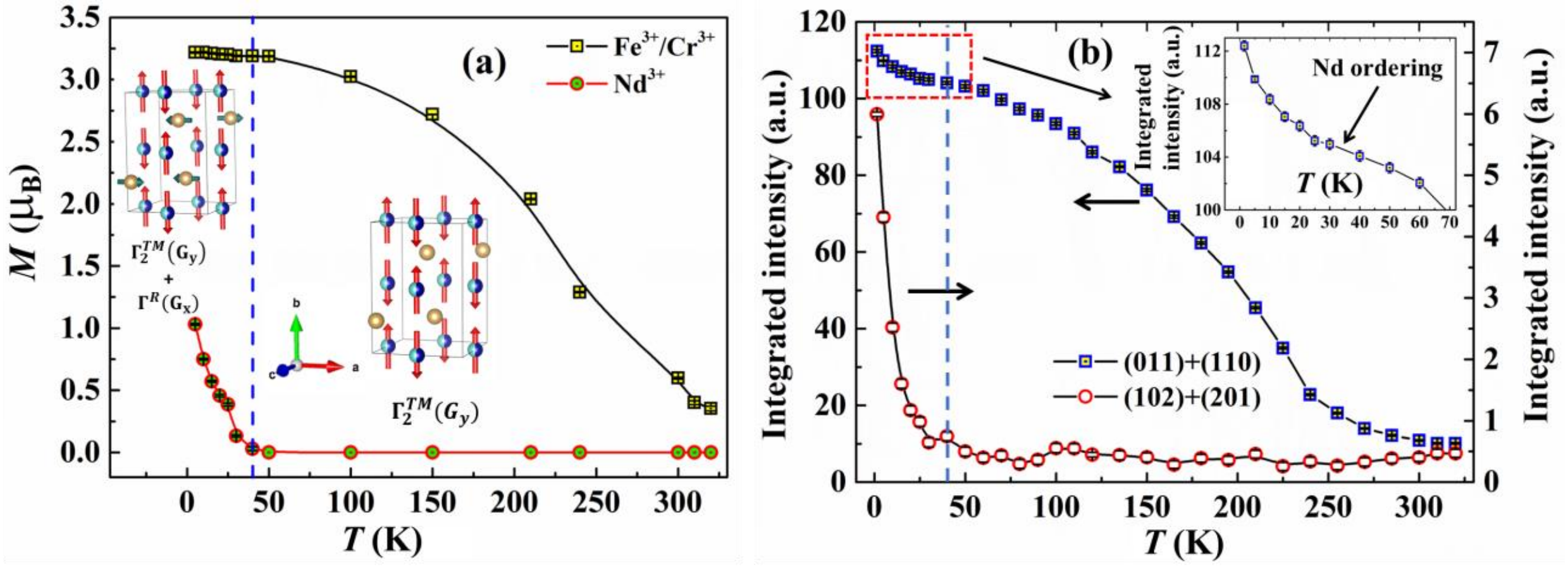

Figure 9: (a) Temperature dependent distributions of magnetic moments for Fe/Cr and Nd sublattices along with irreducible magnetic representation following Bertaut notation. (b) Integrated intensities of (011)+(110) and (102)+(201) magnetic reflections as a function of temperature. The vertical line (inside each data point) attached with each data point corresponds to the error bar of that particular value.

The ordering of Nd can further be justified with attempt of fitting in all physically possible irreducible representations other than $\Gamma_2$ (such as $\Gamma_{1,}$ $\Gamma_{4,}$ $\Gamma_1 + \Gamma_2$ and $\Gamma_2 + \Gamma_4$) at 1.5 K and 15 K as shown in Figure 5(a) and (b) of SM. Any of these representations unable to fit the magnetic Bragg peaks (Figure 5 in SM). However, $\Gamma_2$ representation along with moments at Nd site when considered at 1.5 K, 5 K, 10 K, 15 K, 20 K and 25 K the diffraction pattern fitted excellently including all the magnetic Bragg peaks (Figure 5(a)[v], (b)[v] and (c) in SM). Moreover, only the $\Gamma_2$ representation at the Nd site will give a weak FM mode in the same direction as the one generated by the $Fe^{3+}$/$Cr^{3+}$ site ordering.

Meanwhile, it is important to note the range of SRT for both parent compounds while addressing the issue of strange spin evolution and absence of SRT. SRT in $NdFeO_3$ occurs continuously around 100-170 K [21, 22] while it is a sharp transition for $NdCrO_3$ around 34 K [3, 4]. So, starting from room temperature, it is quite justified in assuming the dominance of Cr over Fe sublattice and the system follow the configuration of Cr i.e, $\Gamma_2$. Now, Fe sublattice in $NdFeO_3$ starts to prefer $\Gamma_2$ ordering around 170 K. It is important to notice that in this temperature range (T < 224 K) Cr sublattice also possesses the same $\Gamma_2$ configuration. Actually, this whole scenario has been described in in Group E. So, there is no ambiguity that the mixed perovskite compound $NdFe_{0.5}Cr_{0.5}O_3$ can be found in $\Gamma_2$ phase in the temperature range of 34-170 K; further dominance of Cr over Fe sublattice is the cause for $NdFe_{0.5}Cr_{0.5}O_3$ to be in the $\Gamma_2$ phase in the temperature range 170-224 K. Now, considering both parent compounds ($NdCrO_3$ and $NdFeO_3$) this has been found from experimental as well as theoretical calculations [44] that Nd-Cr hybridization is stronger than Nd-Fe hybridization. This results a stronger interaction strength of Nd-Cr than Nd-Fe. So, in the temperature range 224 K < T < 245 K, dominant Nd-Cr interaction align the system in $\Gamma_2$ configuration. Above 244 K the system starts entering into the paramagnetic regime, but due to random distribution of Fe/Cr rich clusters neutron diffraction suggests some magnetic contribution from transition metal sublattice. But, the magnitude of Nd-Cr interaction is nearly 2.6 times higher than that of Nd-Fe interaction [44], so it is quite expected that the system would possess $\Gamma_2$ phase only. Another issue is the absence of $\Gamma_1$ phase in $NdFe_{0.5}Cr_{0.5}O_3$ in the vicinity of SRT temperature of $NdCrO_3$ (~34 K), which needs to be addressed. SRT in $NdCrO_3$ has been nicely discussed recently considering the crucial role of Nd-Cr interaction along with the single ion anisotropy (SIA) contribution from Nd and Cr ions [45]. A parameter, $\gamma$ that describes relative strength of Nd-Cr exchange with respect to Cr-Cr exchange was considered as a key factor to analyze the SRT. A suitable range of $\gamma$ (~ 0.35 - 0.62) was calculated for which it is possible to obtain the SRT phase in $NdCrO_3$. When, $\gamma < 0.35$, it favours Cr sublattice to be in $\Gamma_2$ phase rather than experiencing a SRT ($\Gamma_2$ to $\Gamma_1$), while $\gamma > 0.62$ introduces SRT in Nd sublattice rather than in Cr sublattice. So, the effect of *TM-R* interaction, which is considered to be one of the important driving factors for SRT, is stronger in $NdCrO_3$ than $NdFeO_3$. Now, $\gamma$ which is very much dependent on the relative strength of Nd-Cr and Cr-Cr interactions could suffer a significant change in value for $NdFe_{0.5}Cr_{0.5}O_3$ due to the presence of both Cr and Fe in equal proportion. Dilution of Cr sublattice by Fe pushes $\gamma$ away from the critical limit for which SRT in $NdFe_{0.5}Cr_{0.5}O_3$ is not favourable and this results in vanishing of SRT ($\Gamma_2$ to $\Gamma_1$) phase. Thus, low temperature spin orientation of Fe/Cr sublattice follows only $\Gamma_2$ configuration along with ordering of Nd sublattice below 40 K.

**C. Raman Spectroscopy Study:**

Temperature-dependent Raman spectra of $NdFe_{0.5}Cr_{0.5}O_3$ were recorded from 11 to 297 K. No extra Raman peak appears within this temperature range, which rules out the possibility of structural phase transition. Orthorhombic $ABO_3$ type perovskite has twenty-four Raman active modes like $LaMnO_3$, $\Gamma_{Raman} = 7A_g + 5B_{1g} + 7B_{2g} + 5B_{3g}$ [46]. Distinct Raman active phonon modes were found within the detection limit of the instrument at 11 K for $NdFe_{0.5}Cr_{0.5}O_3$ and those are shown in Figure 10. The modes have been assigned following the reported modes of $NdCrO_3$ [47-50] and $NdFeO_3$ [51]. Atomic vibrations associated with these identified modes are shown in Table-II of SM [28] and the fittings with Lorentzian profile have been shown in Figure 6 of SM. Figure 11 shows the temperature evolution of Raman modes in the range of 100-800 $cm^{-1}$. It can be noticed from Figure 11 that with the increase in temperature some modes get red or blue shifted while the intensity of a few modes gradually diminishes. The Raman modes below 200 $cm^{-1}$ are attributed to the vibration of heavy *R*-ions. Above 300 $cm^{-1}$ the modes are characterized by the vibration of light oxygen ions, and in the intermediate range contribution comes from both ions.

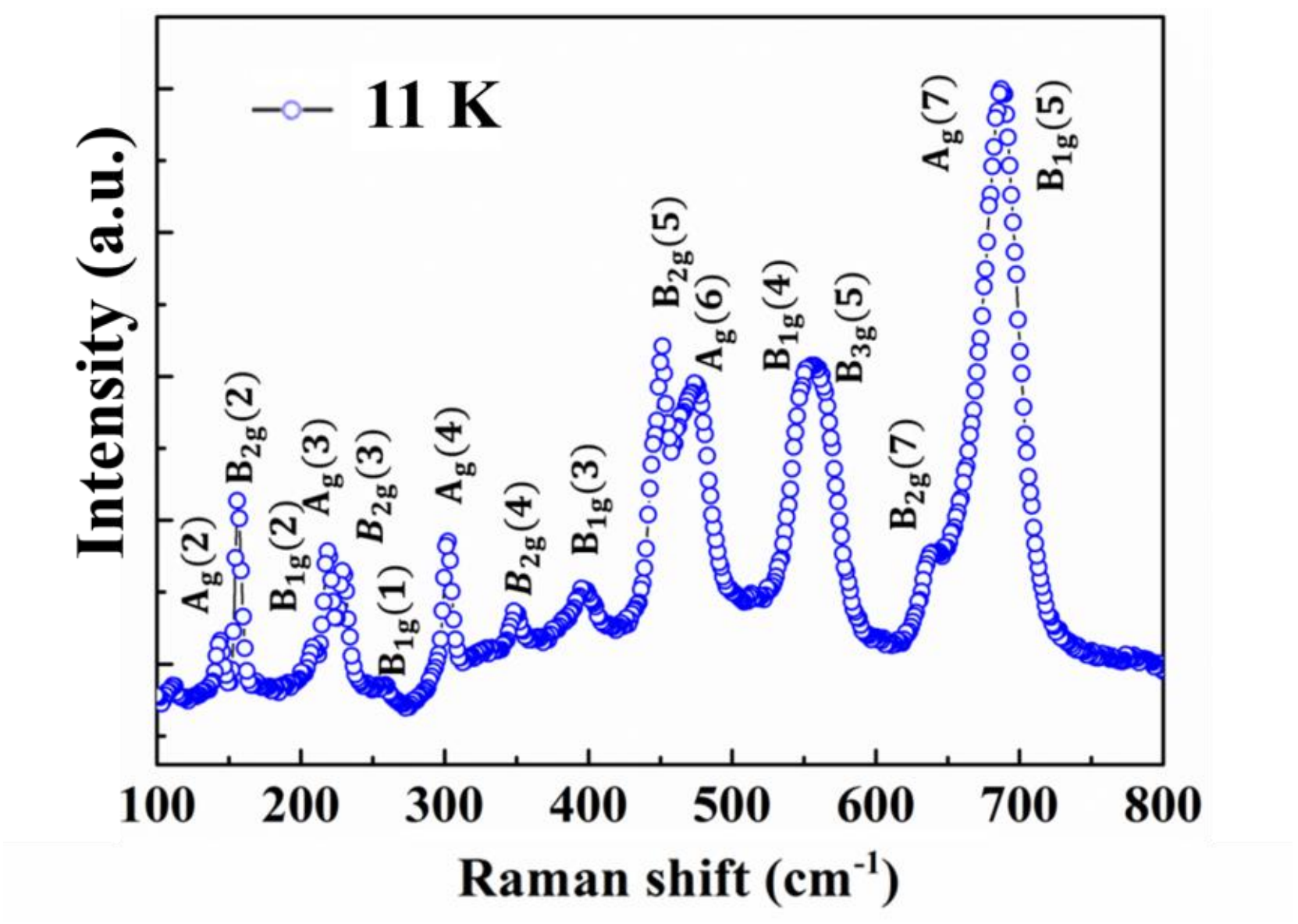

Figure 10: Raman spectra of $NdFe_{0.5}Cr_{0.5}O_3$recorded at 11 K. Assignment of modes is done according to Ref [47-51].

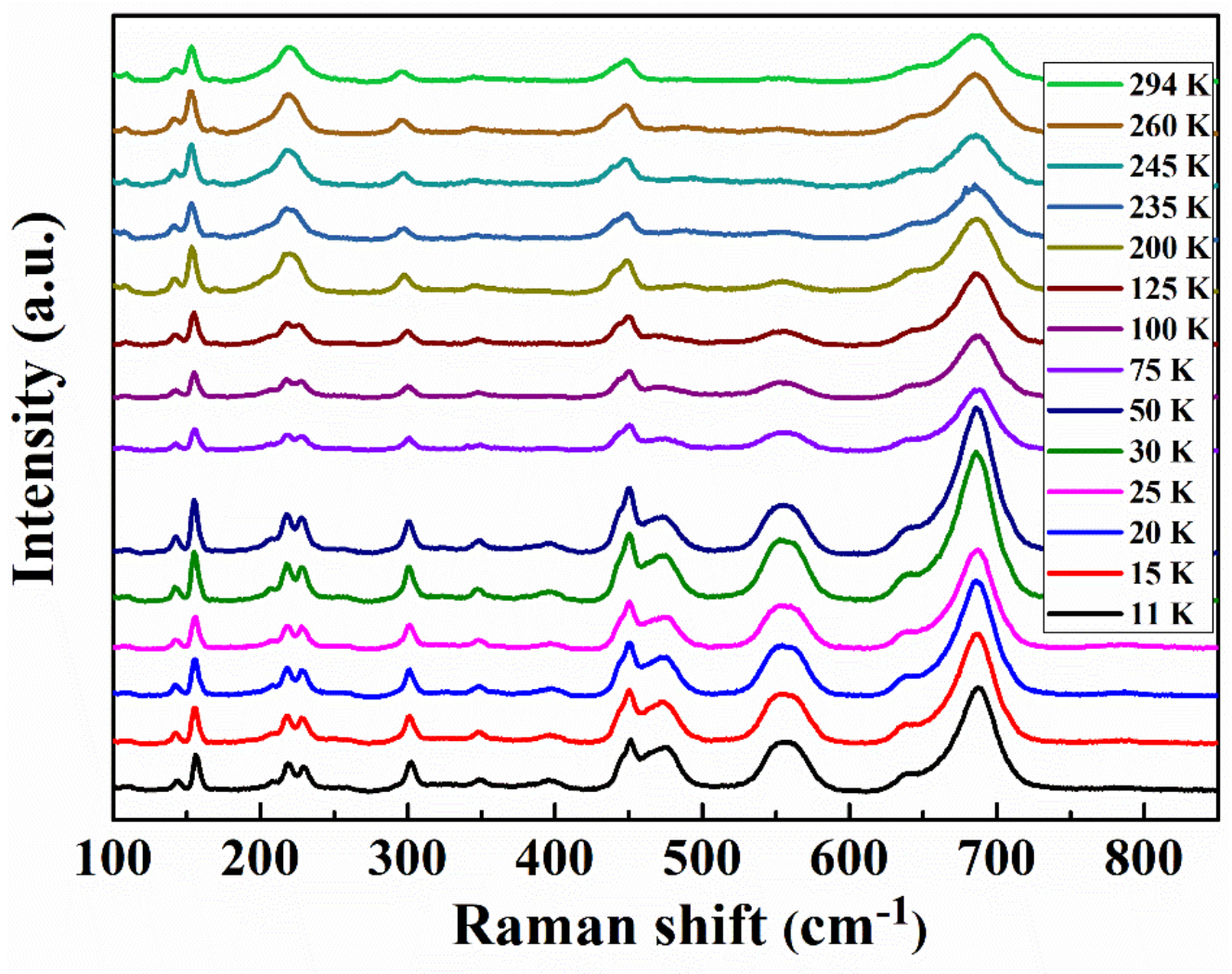

Figure 11: Evolution of Raman modes for $NdFe_{0.5}Cr_{0.5}O_3$ at selective temperatures.

Several reports on $R Fe_{0.5} Cr_{0.5}O_3$ (R= Ho [52], Dy [53], Er and Yb [54]) explained the presence of strong spin-phonon coupling from temperature dependent Raman spectra, which was evident from the anomalous changes in mode frequencies around magnetic transitions. The deviation of mode frequencies from anharmonic behavior and change in phonon lifetime indicate ordering of magnetic sublattice or SRT for some selective modes associated with specific motions of *R* and Fe/Cr ions. In general, temperature dependent wavenumber behavior of a phonon mode in magnetic material follows the equation [55],

$$\omega(T) - \omega(0) = \Delta\omega_{latt} + \Delta\omega_{ren} + \Delta\omega_{anh} + \Delta\omega_{sp-ph} \quad ......(5)$$

where $\omega(0)$ is the phonon wavenumber at 0 K. $\Delta\omega_{latt}$ and $\Delta\omega_{ren}$ are the contributions from lattice volume change with temperature and renormalization of electronic states near magnetic ordering temperature respectively. The first term which deals with isotropic volume change is not important here and for low carrier density the second term can also be ignored. Now, $\Delta\omega_{anh}$ accounts for the wavenumber shift due to anharmonic vibrational potential at constant volume, while $\Delta\omega_{sp-ph}$ describes the spin-phonon coupling present in the system. This anharmonic behavior of phonon modes can be fitted with the following equation as proposed by Balkanski et al. [56]:

$$\omega_{anh}(T) = \omega_0 - C\left(1 + \frac{2}{e^{\frac{\hbar\omega}{2kT}} - 1}\right) - D\left(1 + \frac{3}{e^{\frac{\hbar\omega}{2kT}} - 1} + \frac{3}{\left(e^{\frac{\hbar\omega}{2kT}} - 1\right)^2}\right) \ldots\ldots(6)$$

where $\omega_0$ and C are adjustable parameters. Any deviation from this anharmonic behavior around magnetic transitions is a signature of direct spin-phonon coupling or magnetostriction due to magnetic ordering. Now, presence of spin-phonon coupling changes the lifetime of phonons which can be realized from the changes in full width at half maxima (FWHM) of a particular mode while magnetostriction does not affect phonon lifetime [57]. Few selective Raman modes are considered which are either associated with *R* vibration or motion of Fe/Cr octahedra. One such Raman modes viz. $A_g(4)$ is associated with vibration of Nd and O atoms while another Raman mode $A_g(6)$ appears due to the rotation of Fe/$CrO_6$. Both of these modes show frequency hardening and deviated significantly from anharmonic behavior at low temperature (< 40 K) as evidenced in Figure 12(a) and (b). Also, a deviation can be noticed around Neel temperature of the sample (244 K). Further, the FWHM for $A_g(4)$ modes remains nearly constant at low temperature and then gradually increases as depicted in Figure 12(c). The anomalous behavior of FWHM which is a measure of phonon lifetime, indicates that in the low temperature regime (< 40 K) presence of spin-phonon coupling is strong. In fact the significant change observed in the phonon lifetime of $A_g(4)$ mode below 40 K directs towards the strong spin phonon coupling that indeed reflects the appearance of new ordered magnetic contribution. This spin-phonon coupling can arise from the ordering of *R* sublattice which is an interesting finding. It is observed from NPD study that the *TM* sublattice ordered only in $\Gamma_2$ phase throughout the temperature range of 1.5-320 K and no signature of SRT was observed. However, the ordering of *R* sublattice was prominent below 40 K. So, the deviation of phonon frequency and phonon lifetime in low temperature range (< 40 K) can be attributed to the spin-phonon coupling which arises due to the ordering of *R* sublattice. Hence the observation of *R* ordering below 40 K arising from NPD analysis corroborates the findings of Raman spectroscopy.

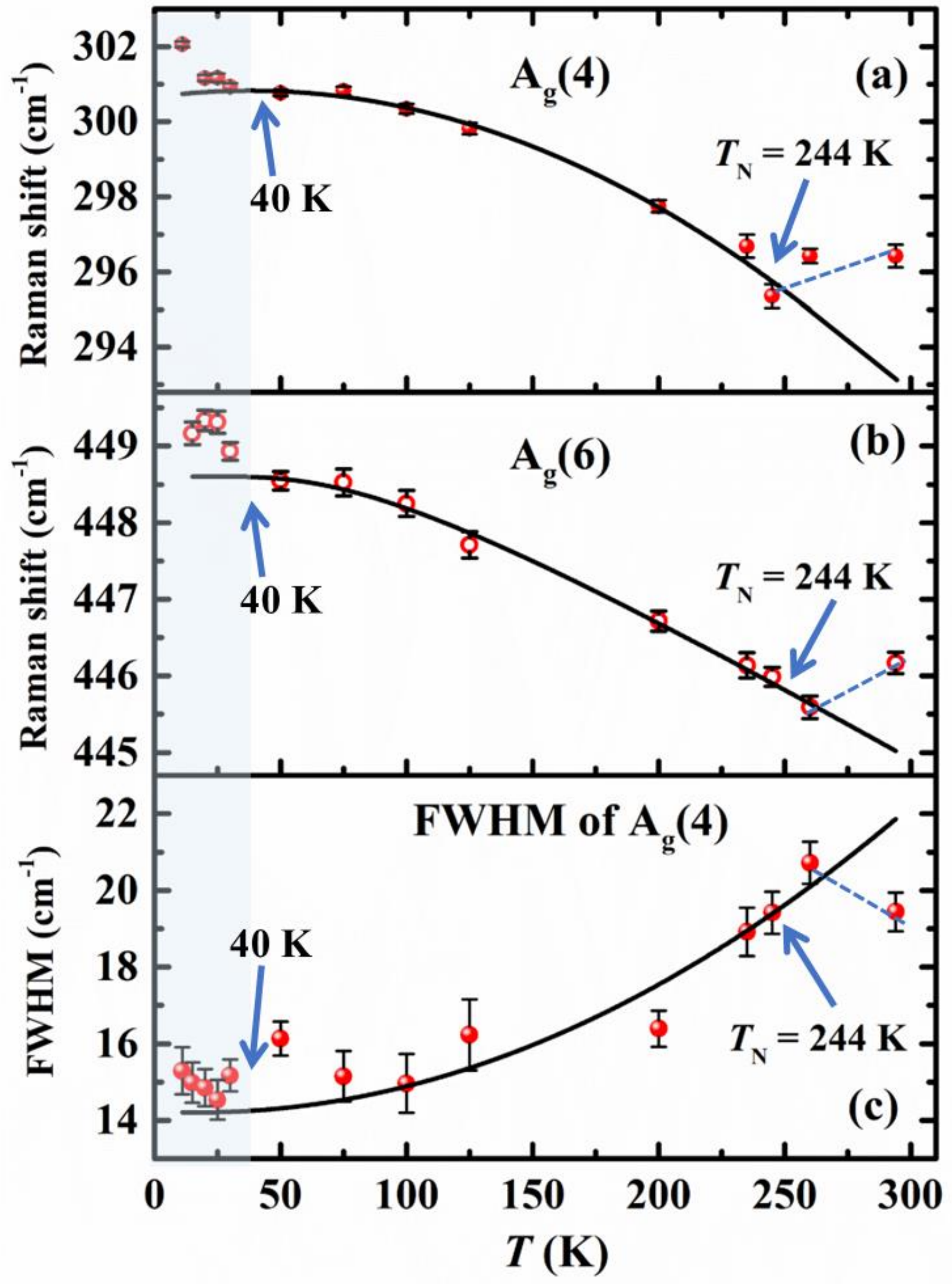

Figure 12(a)-(c): Evolution of wavenumber for $A_g(4)$, $A_g(6)$ modes and respective FWHM of $A_g(4)$ with temperature. The vertical line attached with each data point corresponds to the error bar of that particular value. Black line is the fitting of data points with equation 6. Blue dashed line is used to mark the deviation of experimental data from the fitted curve after $T_N$. The shaded region below 40 K also refers to the deviation which is attributed as ordering of Nd.

**IV. Conclusions**

The synthesized orthochromite-orthoferrite mixed compound, $NdFe_{0.5}Cr_{0.5}O_3$, studied in terms of thorough magnetic characterizations, NPD and Raman spectroscopy, provide interesting findings. The compound exhibits high $T_N$ and the relative orientation of two magnetic sublattices is responsible for negative magnetization. Combined effect of externally applied field and the local induced field at *R* site are responsible for the observed temperature and field dependent relative magnetic alignment as described in the phase diagram. The AFM ordering of the compound provides an advantage over other FM systems, as low to modest magnetic field sweeping is sufficient to obtain BMS which could be utilized in thermomagnetic switch-based device applications. The compound exhibits an appreciable magnetocaloric effect also. The observed BMS and magnetocaloric effect put together suggests $NdFe_{0.5}Cr_{0.5}O_3$ to be a potential candidate for spintronics applications.

Though both parent compounds exhibit spin reorientation transitions where the underlying *TM* sublattices experience a change in easy axis of magnetization, no such phenomenon was observed from NPD studies of $NdFe_{0.5}Cr_{0.5}O_3$ in agreement with findings of an earlier study [18]. NPD study suggests the spin ordering of $Fe^{3+}$/ $Cr^{3+}$ can be assigned to $\Gamma_2$ over the entire temperature range of 1.5-320 K. It is only our study which reflects strong influence of Cr ions in governing the overall spin texture of the compound and the ordering of Fe-rich clusters is responsible for the observed moment above $T_N$ and this observation is consistent with the existence of magnetic peaks even up to 320 K in NPD patterns. Further, in contrast to an earlier study [18], $Nd^{3+}$ in low temperature regime (below 40 K) shows ordering in $C_x$ configuration. This finding, however, is in line with the high magnetic ordering temperature of *R* spins in mixed Tm and Ho compounds [16, 17]. Temperature-dependent Raman spectroscopy study reveals the presence of strong spin-phonon coupling below $T_N$ and also below 40 K corroborating the ordering of Nd sublattice.

**Acknowledgements**

We thank UGC DAE CSR, Indore (CRS-M-292) for financial support and fellowship to S Kanthal; RRCAT, Indore, India for XRPD measurements utilizing their scanning BL 9 of INDUS 2 Synchrotron Source (2.0 GeV, 100 mA); DST, India for financial support and JNCASR, Bengaluru, India for facilitating the NPD experiments (ISIS Beamtime application No. RB1920201) at Wish Beamline, ISIS neutron source, Rutherford Appleton Laboratory, UK. We acknowledge Mr. A. Upadhyay and Dr. A. K. Sinha of RRCAT, Indore, India for XRPD measurements and Prof. A. Barman of SNNCBS, Kolkata, India for his advice in choosing this research problem.

**Reference**

**[1]** L. T. Tsymbal, Ya. B. Bazaliy, V. N. Derkachenko, V. I. Kamenev, G. N. Kakazei, F. J. Palomares, and P. E. Wigen, J. Appl. Phys. **101**, 123919 (2007).

**[2]** Ya. B. Bazaliy, L. T. Tsymbal, G. N. Kakazei, A. I. Izotov, and P. E. Wigen, Phys. Rev. B **69**, 104429 (2004).

**[3]** N. Shamir, H. Shaked, and S. Shtrikman, Phys. Rev. B **24**, 6642 (1981).

**[4]** Eric Bousquet and Andrés Cano, J. Phys.: Condens. Matter **28**, 123001 (2016) (28pp).

**[5]** L. Bellaiche, G. Zhigang, and A. K. Igor, J. Phys.: Cond. Matter. **24**, 312201 (2012).

**[6]** E. F. Bertaut, Acta Crystallographica Section A **24**, 217 (1968).

**[7]** Y. Su, J. Zhang, Z. Feng, L. Li, B. Li, Y. Zhou, Z. Chen, and S. Cao, J. Appl. Phys. **108**, 013905 (2010).

**[8]** K. Yoshii, J. Solid State Chem. **159**, 204 (2001).

**[9]** J. Mao, Y. Sui, X. Zhang, Y. Su, X. Wang, Z. Liu, Y. Wang, R. Zhu, Y.Wang,W. Liu, and J. Tang, Appl. Phys. Lett. **98**, 192510 (2011).

**[10]** P. Mandal, A. Sundarean, C. N. R. Rao, A. Lyo, P. M. Shirage, Y. Tanaka, Ch. Simon, V. Pralong, O. I. Lebedev, V. Caignaert, and B. Raveau, Phys. Rev. B **82**, 100416(R) (2010).

**[11]** T. Bora and S. Ravi, J. Appl. Phys. **114**, 033906 (2013).

**[12]** D. Delmonte, F. Mezzadri, C. Pernechele, G. Calestani, G. Spina, M. Lantieri, M. Solzi, R. Cabassi, F. Bolzoni, A. Migliori, C. Ritter, and E. Gilioli**,** Phys. Rev. B **88**, 014431 (2013).

**[13]** T. Bora and S. Ravi**,** J. Magn. Magn. Mater. **386**, 85 (2015).

**[14]** A. K. Azad, A. Mellergȧrd, S.-G. Eriksson, S. A. Ivanov, S. M. Yunus, F. Lindberg, G. Svensson, and R. Mathieu, Mater. Res. Bull. **40**, 1633 (2005).

**[15]** E. Bousquet and A. Cano, J. of Phys.: Condensed Matter **28**, 123001 (2016).

**[16]** F. Pomiro, R. D. Sánchez, G. Cuello, A. Maignan, C. Martin, and R. E. Carbonio, Phys. Rev. B **94**, 134402 (2016).

**[17]** J. P. Bolletta, F. Pomiro, R. D. Sánchez, V. Pomjakushin, G. Aurelio, A. Maignan, C. Martin, and R. E. Carbonio, Phys. Rev. B **98**, 134417 (2018).

**[18]** M. P. Sharannia, S. De, R.Singh, A. Das, R. Nirmala, and P.N. Santhosh, J. Magn. Magn. Mater, **430**, 109 (2017).

[**19**] S. Lei, L. Liu, C. Wang, C. Wang, D. Guo, S. Zeng, B. Cheng, Y. Xiao, and L. Zhou, J. Mater. Chem. A **1,** 11982 (2013).

**[20]** F. Bartolome, J. Bartolome, M. Castro, and J. J. Melero, Phys. Rev. B **62,** 1058 (2000).

**[21]** S. J. Yuan, Y. M. Cao, L. Li, T. F. Qi, S. X. Cao, J. C. Zhang, L. E. DeLong, and G. Cao, J. Appl. Phys. **114,** 113909 (2013).

**[22]** E. Constable, D. L. Cortie, J. Horvat, R. A. Lewis, Z. Cheng, G. Deng, S. Cao, S. Yuan, and G. Ma, Phys. Rev. B **90,** 054413 (2014).

**[23]** W. Sławiński, R. Przeniosło, I. Sosnowska, and E. Suard, J. Phys.: Condens. Matter **17,** 4605 (2005).

**[24]** L. C. Chapon *et al.*, Neutron News **22**, 22 (2011).

**[25]** H. Rietveld, J. Appl. Crystallogr. **2**, 65 (1969).

**[26]** J. Rodríguez-Carvajal, Physica B: Condensed Matter. **192**, 55 (1993).

**[27]** V. Petrícek, M. Dusek and L. Palatinus, Zeitschrift für Kristallographie - Crystalline Materials **229,** 345 (2014).

**[28]** See Supplemental Material for additional information in crystal structure analysis from powder XRD, negative magnetization in M-T, temperature dependent NPD, and Raman spectroscopy analysis of $NdFe_{0.5}Cr_{0.5}O_3$.

**[29]** L. H. Yin, J. Yang, P. Tong, X. Luo, W. H. Song, J. M. Dai, X. B. Zhu, and Y. P. Sun, Appl. Phys. Lett. **110**, 192904 (2017).

**[30]** N. Dasari, P. Mandal, A. Sundaresan, and N. S. Vidhyadhiraja, Euro. Phys. Lett, **99**, 17008 (2012).

**[31]** S. J. Yuan, W. Ren, F. Hong, Y. B. Wang, J. C. Zhang, L. Bellaiche, S. X. Cao, and G. Cao, Phys. Rev. B **87**, 184405 (2013).

**[32]** C. Li, T. Yan, C. Chakrabarti, R. Zhang, X. Chen, Q. Fu, S. Yuan, and G. O. Barasa, J. Appl. Phys. **123**, 093902 (2018).

**[33]** N. Kumar, and A. Sundaresan, Solid State Commun. **150**, 1162 (2010).

**[34]** P. S. R. Murthy, K. R. Priolkar, P. A. Bhobe, A. Das, P. R. Sarode, and A. K. Nigam, J. Magn. Magn. Mater. **322**, 3704 (2010).

**[35]** R. P. Singh, and C. V. Tomy, Phys. Rev. B **78**, 024432 (2008).

**[36]** J. Y. Yang, X. D. Shen, V. Pomjakushin, L. Keller, E. Pomjakushina, Y. W. Long, and M. Kenzelmann, Phys. Rev. B **101**, 014415 (2020).

**[37]** S. K. Neogi, S. Chattopadhyay, R. Karmakar, A. Banerjee, S. Bandyopadhyay, and A. Banerjee, J. Alloys conpd. 573, 76 (2013).

**[38]** X. Liu, L. Hao, Y. Liu, X. Ma, S. Meng, Y. Li, J. Gao, H. Guo, W. Han, K. Sun et al., J. Magn. Magn. Mater. **417**, 382 (2016).

**[39]** B. Tiwari, M. K. Surendra, and M. S. R. Rao, J. Phys.: Condens. Matter **25** 216004 (2013)

**[40]** Y. Su, J. Zhang, Z. Feng, Z. Li, Y. Shen, and S. Cao, J. Rare Earths **29** 1060 (2011)

**[41]** S. Yuan, Y. Yang, Y. Cao, A. Wu, B. Lu, S. Cao, and J. Zhang, Solid State Commun. **188** 19 (2014).

**[42]** M. Shao, S. Cao, Y. Wang, S. Yuan, B. Kang, J. Zhang, A. Wu, and J. Xu, J. Cryst. Growth **318** 947 (2011).

**[43]** A. Bombik, B. Leśniewska, J. Mayer, A. Oleś, A. W. Pacyna and J. Przewoźnik, J. Magn. Magn. Mater. **168** 139 (1997).

**[44]** F. Bartolomé, J. Bartolomé, M. Castro, and J. J. Melero, Phys. Rev. B **62**, 1058 (2000).

**[45]** H. Das, A. F. Rébola, and T. Saha-Dasgupta, Phys. Rev. Mater. **5**, 124416 (2021).

**[46]** M. N. Iliev, M. V. Abrashev, H.-G. Lee, V. N. Popov, Y. Y. Sun, C. Thomsen, R. L. Meng, and C. W. Chu, Phys. Rev. B **57**, 2872 (1998).

**[47]** N. R. Camara, V. T. Phuoc, I. M. Laffez, and M. Zaghriou, J Raman Spectrosc. **48**, 1839 (2017).

**[48]** S. Saha, S. Chanda, A. Dutta, and T. P. Sinha, J Sol-Gel Sci Technol. **69**, 553 (2014).

**[49]** M. C. Weber, J. Kreisel, P. A. Thomas, M. Newton, K. Sardar, and R. I. Walton, Phys. Rev. B 85, 054303 (2012).

**[50]** K. D. Singh, R. Pandit, and R. Kumar, Solid State Sci. **85**, 70 (2018).

**[51]** S. Chanda, S. Saha, A. Dutta, and T.P. Sinha, Mater. Res. Bull. 48, 1688 (2013).

**[52]** G. Kotnana, V. G. Sathe, and S. Narayana Jammalamadak, J. Raman Spectrosc. **49**, 764 (2018).

**[53]** L. H. Yin, J. Yang, R. R. Zhang, J. M. Dai, W. H. Song, and Y. P. Sun, App. Phys. Lett. **104**, 032904 (2014).

**[54]** K. Yadav, G. Kaur, M. K. Sharma, and K. Mukherjee, Phys. Lett. A **384**, 126638 (2020).

**[55]** E. Granado, A. García, J. A. Sanjurjo, C. Rettori, I. Torriani, F. Prado, R. D. Sánchez, A. Caneiro, and S. B. Oseroff, Phys. Rev. B **60**, 11879 (1999).

**[56]** M. Balkanski, R. F. Wallis, and E. Haro, Phys. Rev. B **28**, 1928 (1983).

**[57]** A. Nonato, B. S. Araujo, A. P. Ayala, A. P. Maciel, S. Yanez Vilar, M. Sanchez Andujar, M. A. Senaris Rodriguez, and C. W. A. Paschoal, Appl. Phys. Lett. **105**, 222902 (2014).

# Supplemental Material

**Crystal structure from XRD:**

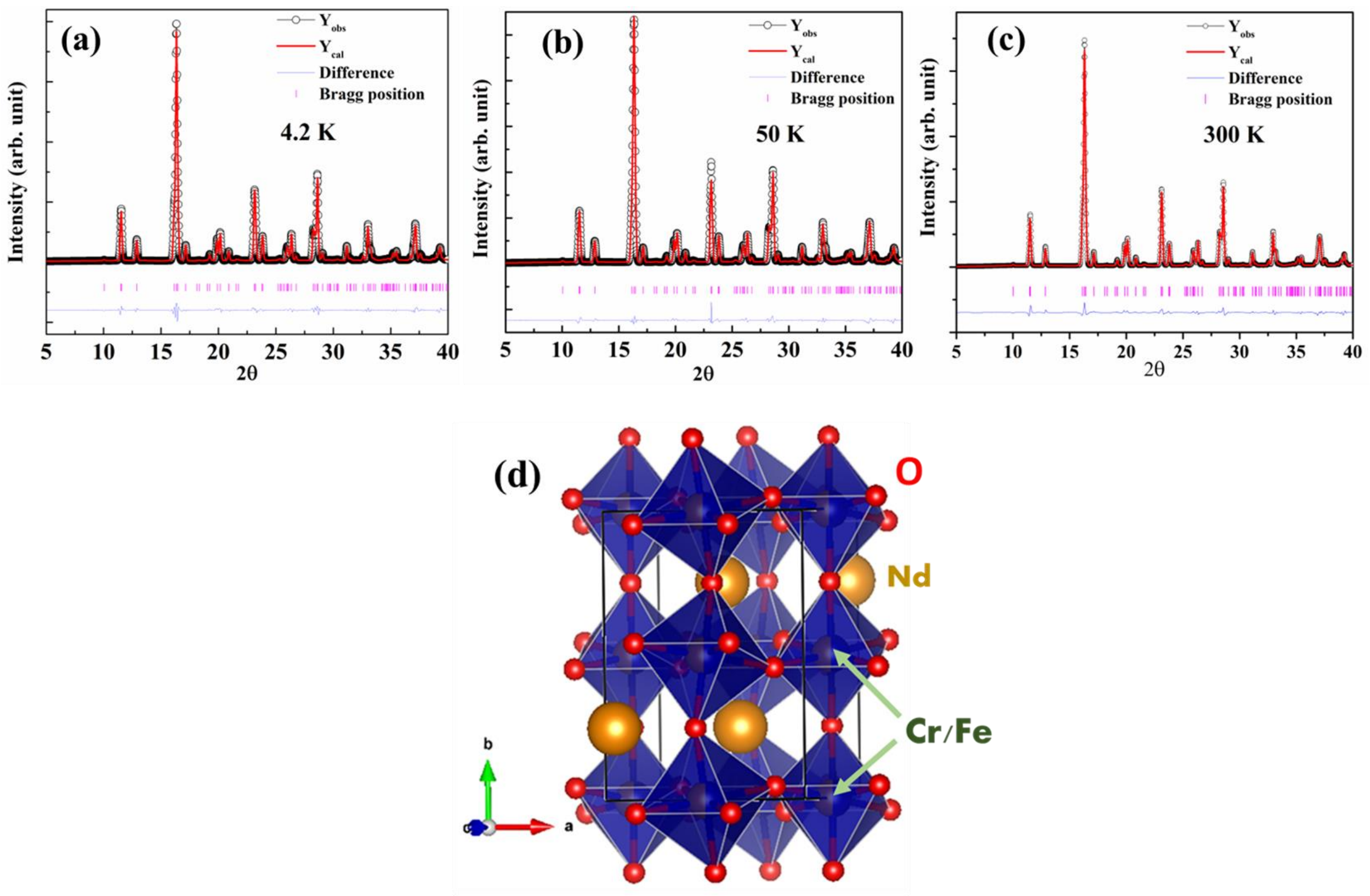

Figure 1(a)-(c): *Rietveld refined XRD patterns at 4.2, 50, and 300 K respectively*. (d) Orthorhombic crystal structure with *Pnma* space group.

Table-I: Lattice parameters of $NdFe_{0.5}Cr_{0.5}O_3$ at some selected temperatures as obtained from powder XRPD and NPD analysis.

| T (K) | a (Å) | | b (Å) | | c (Å) | | V (Å³) | |
|---|---|---|---|---|---|---|---|---|
| | XRPD | NPD | XRPD | NPD | XRPD | NPD | XRPD | NPD |
| 4.2 | 5.54144(4) | | 7.72479(7) | | 5.43431(3) | | 232.624(6) | |
| 5 | - | 5.53340(2) | - | 7.71724(2) | - | 5.42871(4) | - | 231.818(6) |
| 50 | 5.54393(3) | 5.53350(3) | 7.72960(4) | 7.71755(5) | 5.43748(9) | 5.42882(2) | 233.009(7) | 231.837(5) |
| 300 | 5.54486(8) | 5.53562(4) | 7.74188(8) | 7.73194(5) | 5.44640(4) | 5.43880(4) | 233.801(1) | 232.787(3) |

**Negative Magnetization Analysis:**

Overall magnetization in $NdFe_{0.5}Cr_{0.5}O_3$ can be attributed to the contribution from both Cr/Fe and Nd sublattices. Observed negative magnetization under certain low applied field can be explained by the effect of induced field at Nd site due to Cr/Fe sublattice. Following equation successfully explains the magnetic behaviour at negative region.

$$M = M_{Cr^{3+}/Fe^{3+}} + \frac{C(H_I + H_{Ext})}{T - \theta_C} \ldots\ldots (1)$$

The FC curves under different $H_{Ext}$ are fitted with the above equation and value of $M_{Cr^{3+}/Fe^{3+}}$ and $H_I$ are found to be in the range of 0.65±0.01 emu/g to 0.98±0.01 emu/g and -110±0.002 Oe to 1502±0.0001 Oe respectively for $H_{Ext}$=100 Oe to 1500 Oe. It is noteworthy that with increase in the applied field, $M_{Cr^{3+}/Fe^{3+}}$ and $H_I$ increase which can be attributed to the increasing AFM domain size. Fitting of FC curve for 100 Oe field is shown in the figure below, which shows a good agreement with the above model.

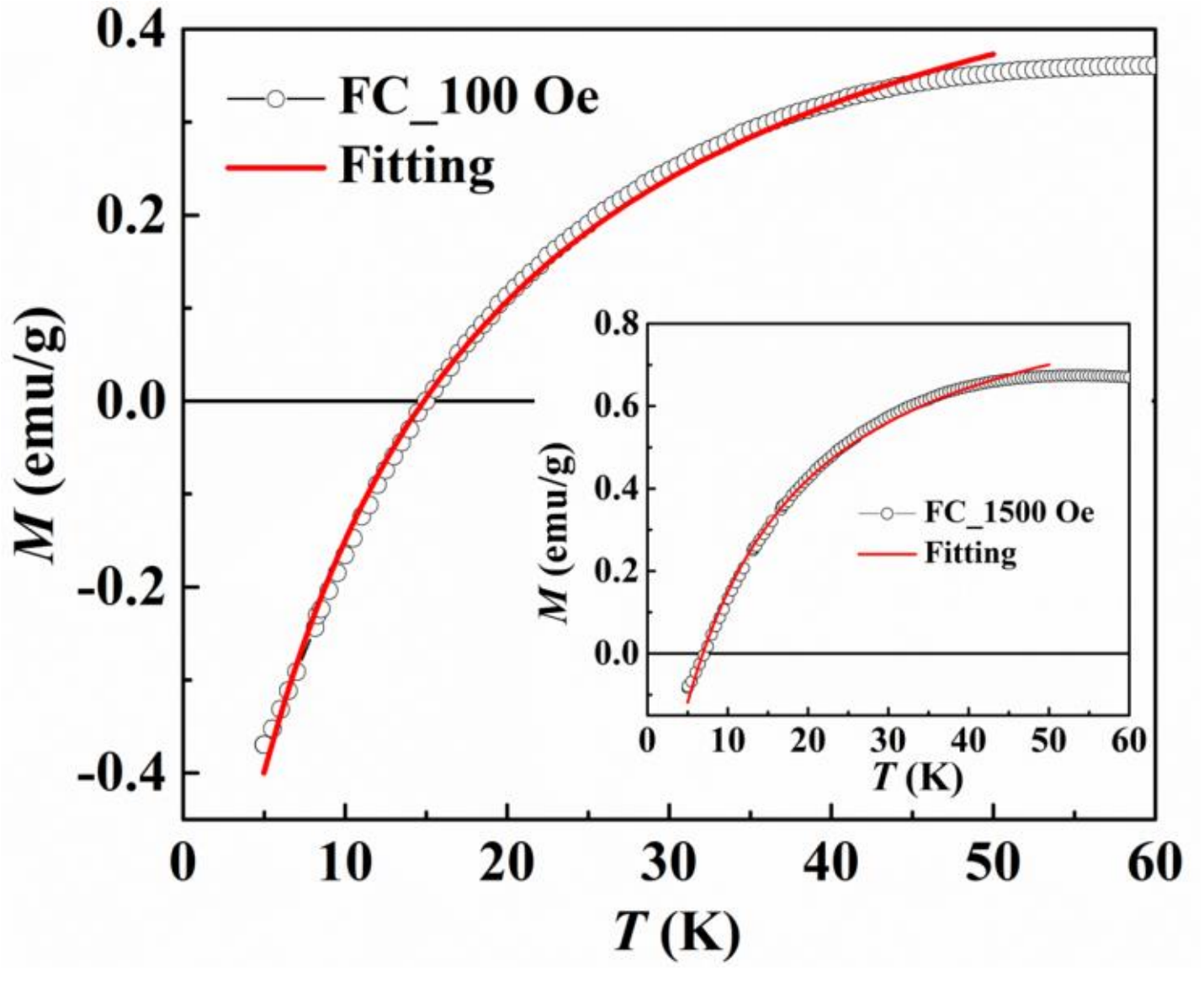

Figure. 2: Fitting of FC curve with equation 1 for 100 Oe and 1500 Oe (inset).

**Non collinear AFM structure analysis at 4b site from Neutron diffraction data**

As temperature is increased from 1.5 K the intensity of both the peaks (110) and (011) which show main magnetic contribution from Fe/Cr sublattice, decreases as shown below.

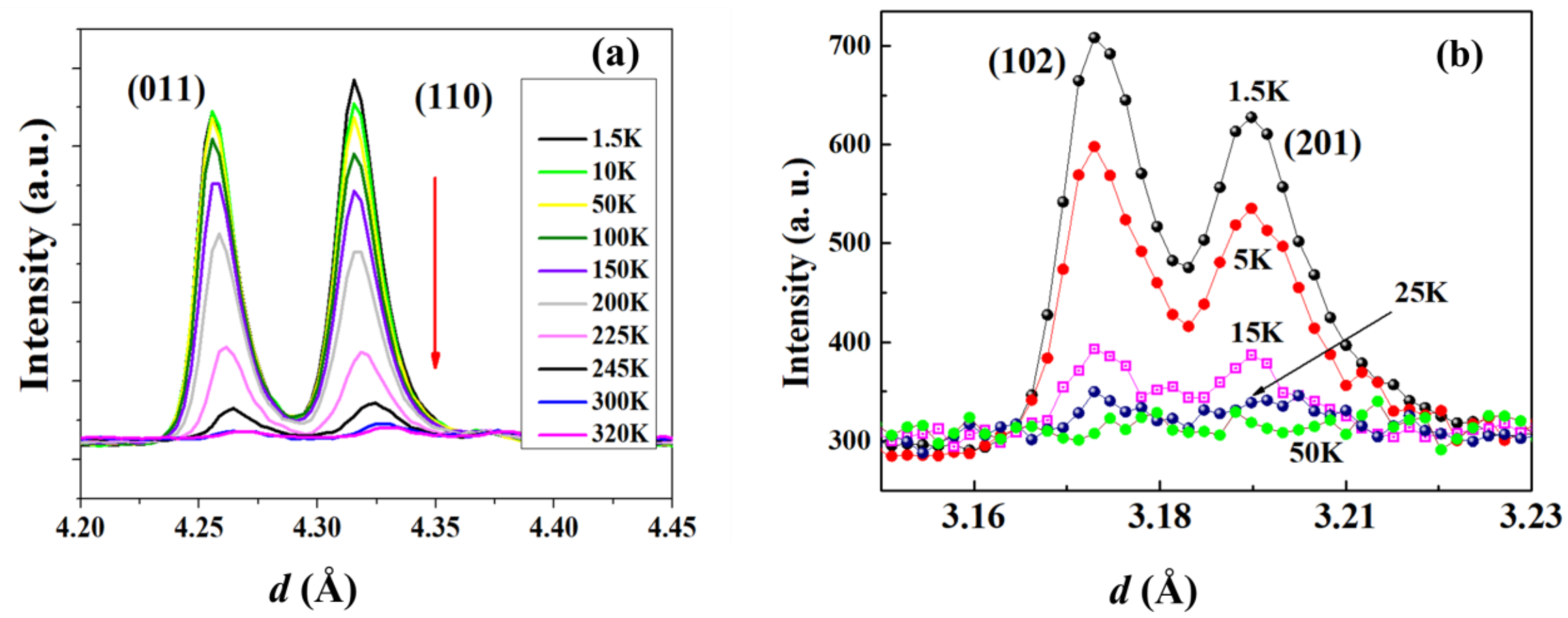

Figure 3: Temperature dependent NPD spectra of $NdFe_{0.5}Cr_{0.5}O_3$ showing (a) (011) and (110), (b) (102) and (201) magnetic peaks.

According to Bertaut notations, four collinear magnetic spins (here $Cr^{3+}$/$Fe^{3+}$) in 4b sites of a unit cell- S1(0, 0, 0.5), S2(0.5, 0, 1), S3(0, 0.5, -0.5), S4(0.5, 0.5, 0) couple to form basic spin ordering states necessary to describe the noncolinear spin texture of perovskites. Depending on the relative directions of four 4b spins in a unit cell four base vectors were introduced as:

$$F = +S1 + S2 + S3 + S4,$$
$$G = +S1 - S2 - S3 + S4,$$
$$C = +S1 - S2 + S3 - S4,$$
$$A = +S1 + S2 - S3 - S4,$$

Bertaut explained all possible irreducible representations for the elements at 4b and 4c sites in orthorhombic centrosymmetric perovskite compounds having $Pbnm(D_{2h}^{16})$ space group. All those 8 possible irreducible representations consistent with *Pnma* space group notation are listed in Table-II. It is found that for an alloy compound having two transition metal sharing 4b site, magnetic phase can be explained by virtue of these irreducible representations mentioned in Table-II.

Table-II: Irreducible representation describing the magnetic structure of space group *Pnma* with wave vector $k$=(0, 0, 0) along with magnetic Shubnikov space group.

| **Irreducible representation** | **Transition metal in 4b site ($Cr^{3+}$/$Fe^{3+}$)** | **Rare earth in 4c ($Nd^{3+}$)** | **Magnetic Space group** |
|---|---|---|---|
| $\Gamma_1$ | $G_x\ C_y\ A_z$ | $C_y$ | *Pnma* |
| $\Gamma_2$ | $C_x\ G_y\ F_z$ | $C_x\ F_z$ | *Pn'm'a* |
| $\Gamma_3$ | $F_x\ A_y\ C_z$ | $F_y\ C_z$ | *Pnm'a'* |
| $\Gamma_4$ | $A_x\ F_y\ G_z$ | $F_y$ | *Pn'ma'* |
| $\Gamma_5$ | - | $A_x\ G_z$ | *Pn'm'a'* |
| $\Gamma_6$ | - | $A_y$ | *Pnma'* |
| $\Gamma_7$ | - | $G_y$ | *Pn'ma* |
| $\Gamma_8$ | - | $G_x\ A_z$ | *Pnm'a* |

NPD studies considering wave vector k=(0,0,0) result G type AFM ordering of Cr/Fe sublattice in $\Gamma_2$ spin configuration over entire temperature range. Fitting of (011) and (110) magnetic peaks with $\Gamma_2$ orientation is presented in Figure 4. A good agreement of experimental and simulated data justifies the proper selection of spin model.

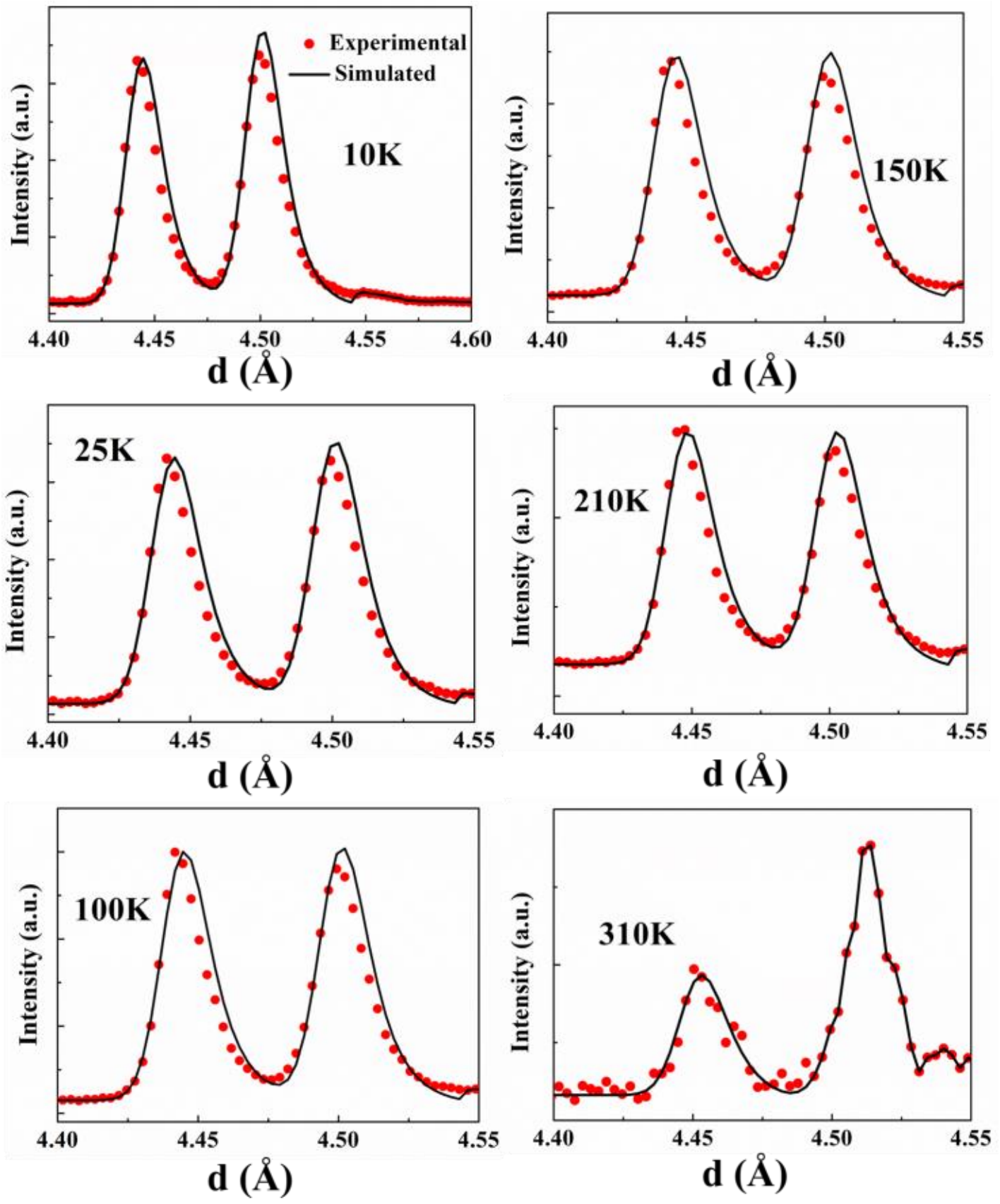

Figure 4: Fitting of magnetic intensities (011 and 110) with $\Gamma_2$ spin model at different temperatures.

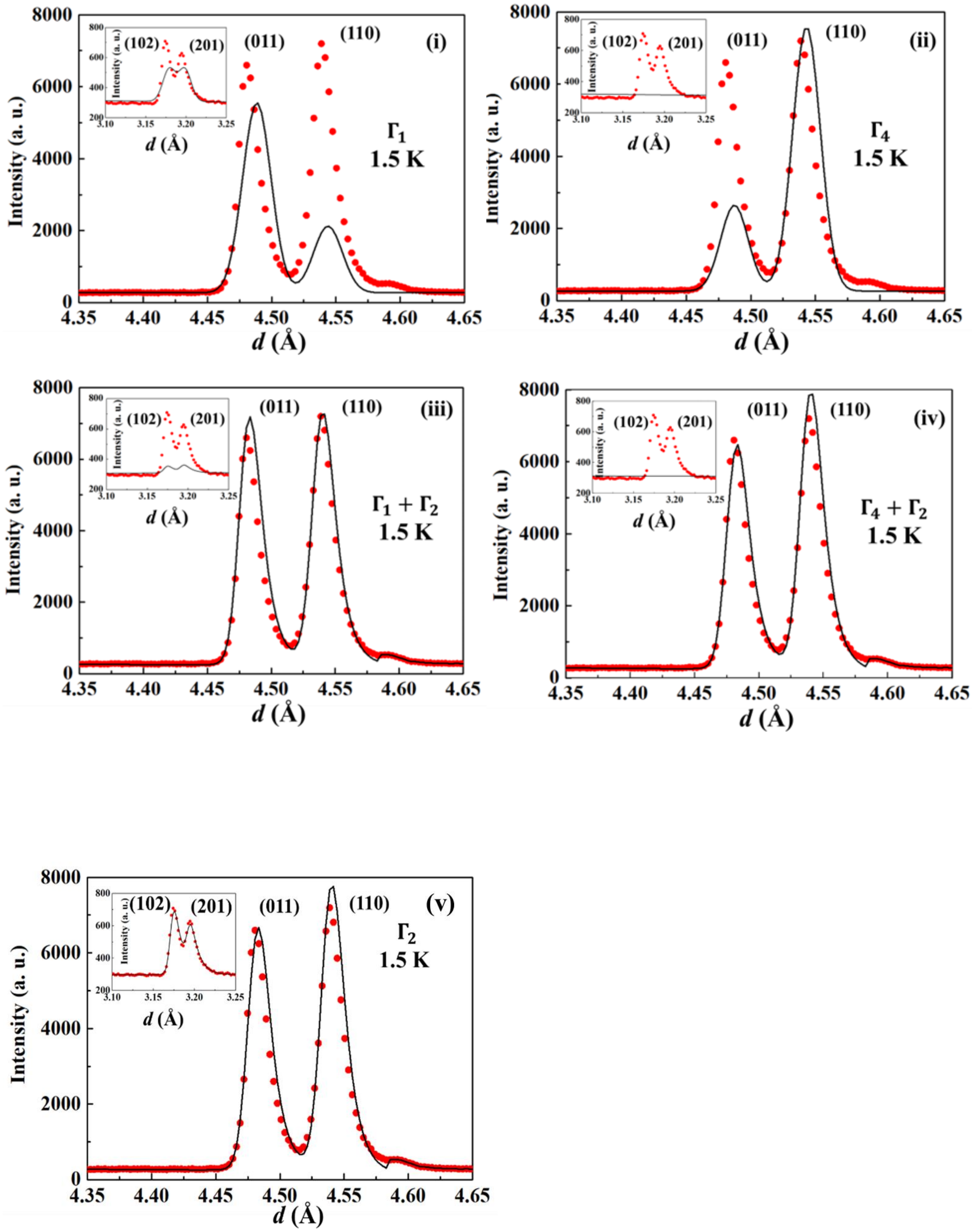

Figure 5 (a): Fittings at 1.5 K with all allowed irreducible representations: (i) $\Gamma_1$, (ii) $\Gamma_4$, (iii) $\Gamma_1 + \Gamma_2$, (iv) $\Gamma_2 + \Gamma_4$ and (v) $\Gamma_2$ with moments at Nd site

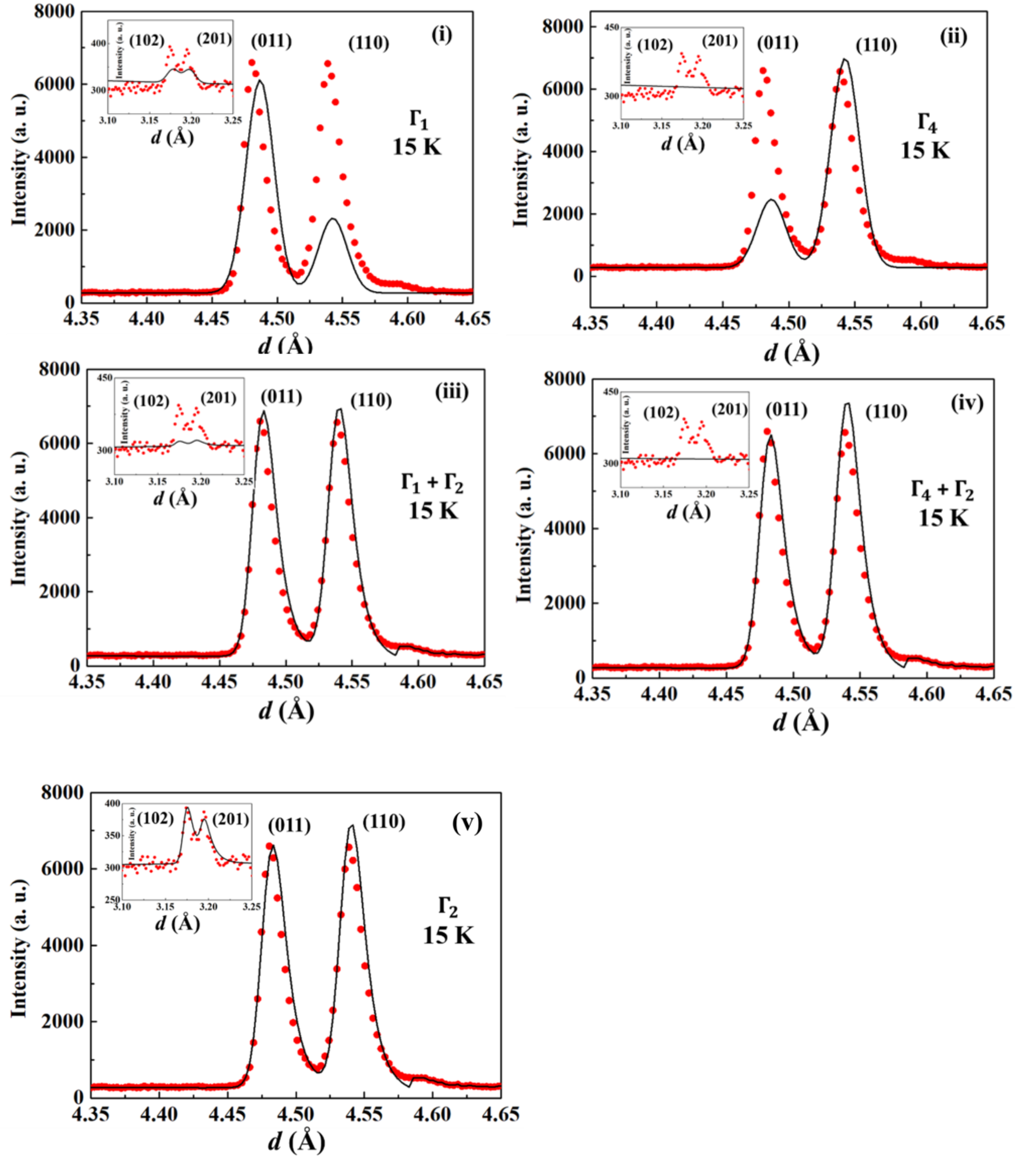

Figure 5 (b): Fittings at 15 K with all allowed irreducible representations: (i) $\Gamma_1$, (ii) $\Gamma_4$, (iii) $\Gamma_1 + \Gamma_2$, (iv) $\Gamma_2 + \Gamma_4$ and (v) $\Gamma_2$ with moments at Nd site

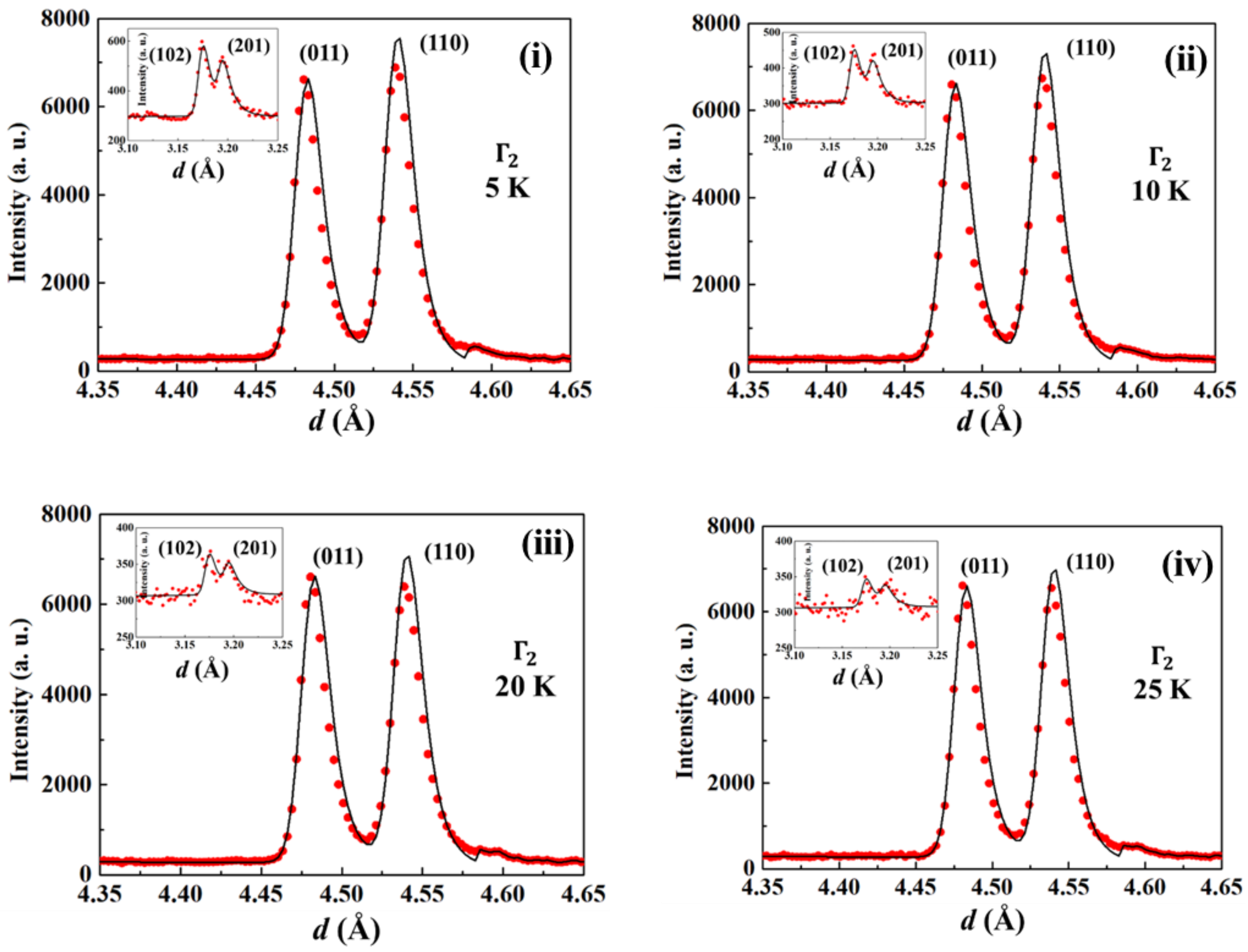

Figure 5 (c): Fittings of $\Gamma_2$ irreducible representations with moments at Nd site some specified temperatures: (i) 5 K, (ii) 10 K, (iii) 20 K (iv) 25 K

In Figure 5 [a (i-iv)] and [b (i-iv)], we have shown fittings of NPD pattern at 1.5 K, 15 K considering $\Gamma_1$, $\Gamma_4$, $\Gamma_1$+ $\Gamma_2$, $\Gamma_2$+ $\Gamma_4$ phases without moments at Nd site. All the curves exhibit poor fitting. The curves for $\Gamma_4$ and $\Gamma_2$+ $\Gamma_4$ even didn't consider the low temperature magnetic Bragg peaks (102) and (201), while the fittings of (102) and (201) peaks for $\Gamma_1$ and $\Gamma_1$+ $\Gamma_2$ phases are also extremely bad. Now we have fitted the 1.5 K and 15 K curves in $\Gamma_2$ phase with moments at the Nd site in Figure 5 [a (v)] and [b (v)] and it shows extremely good fitting. Thereafter, we have provided fittings for magnetic peaks at other low temperatures (5 K, 10 K, 20 K and 25 K) in Figure 5 (c) (i-iv) in $\Gamma_2$ phase with moments at the Nd site and those curves also show excellent fitting. Hence, there is only one way to fit the NPD data in physically meaningful way is to include magnetic moments on Nd site and $\Gamma_2$ phase for Fe/Cr sublattice. Actually, these fittings clearly show the absence of spin reorientation transition and ordering of Nd in the low temperature regime.

**Identification of Raman modes for $NdFe_{0.5}Cr_{0.5}O_3$**

Raman active modes and the main atomic motions behind them for $NdFe_{0.5}Cr_{0.5}O_3$ are assigned [1-5] and listed below in Table-III.

Table-III: Phonon modes as obtained for $NdFe_{0.5}Cr_{0.5}O_3$, their symmetry and main atomic motion associated with them.

| **Raman Modes for** $NdFe_{0.5}Cr_{0.5}O_3$ | **Symmetry** | **Main atomic motion [1, 2, 4 ]** |
|---|---|---|
| | $A_g(1)$ | R(x) O2(x,z) |
| 143 | $A_g(2)$ | R(z) out of plane |
| 218 | $A_g(3)$ | $TMO_6$ in phase y rotations |
| 300 | $A_g(4)$ | O1(x), R(-x) |
| | $A_g(5)$ | $TMO_6$ out-of-phase x rotations |
| 472 | $A_g(6)$ | Out-of-phase $TMO_6$ bending |
| 678 | $A_g(7)$ | In-phase stretching of $TMO_6$ |
| 254 | $B_{1g}(1)$ | R(y) |
| 208 | $B_{1g}(2)$ | Out-of-phase rotations of $CrO_6$ octahedra around y |
| 395 | $B_{1g}(3)$ | In-phase rotations of $CrO_6$ octa-hedra around x |
| 551 | $B_{1g}(4)$ | Out-of-phase stretching of $CrO_6$ octahedra around xz-planes |
| 688 | $B_{1g}(5)$ | Out-of-phase stretching of $CrO_6$ octahedra around y |
| | $B_{2g}(1)$ | R(z) |
| 156 | $B_{2g}(2)$ | R(x) |
| 229 | $B_{2g}(3)$ | Out-of-phase rotations of $CrO_6$ octahedra around z |
| 350 | $B_{2g}(4)$ | OI(z) |
| 450 | $B_{2g}(5)$ | Out-of-phase bending of $CrO_6$ octrahedra |
| | $B_{2g}(6)$ | In-phase bending of $CrO_6$ octra-hedra |
| 639 | $B_{2g}(7)$ | In-phase stretching of $CrO_6$ oc-trahedra around xz-planes |
| | $B_{3g}(1)$ | R(y) |
| | $B_{3g}(2)$ | In-phase rotations of $CrO_6$ octa-hedra around z |
| | $B_{3g}(3)$ | Out-of-phase bending of $CrO_6$ octrahedra |
| | $B_{3g}(4)$ | Out-of-phase stretching of $CrO_6$ octrahedra |
| 565 | $B_{3g}(5)$ | Out-of-phase breathing of $CrO_6$ octrahedra |

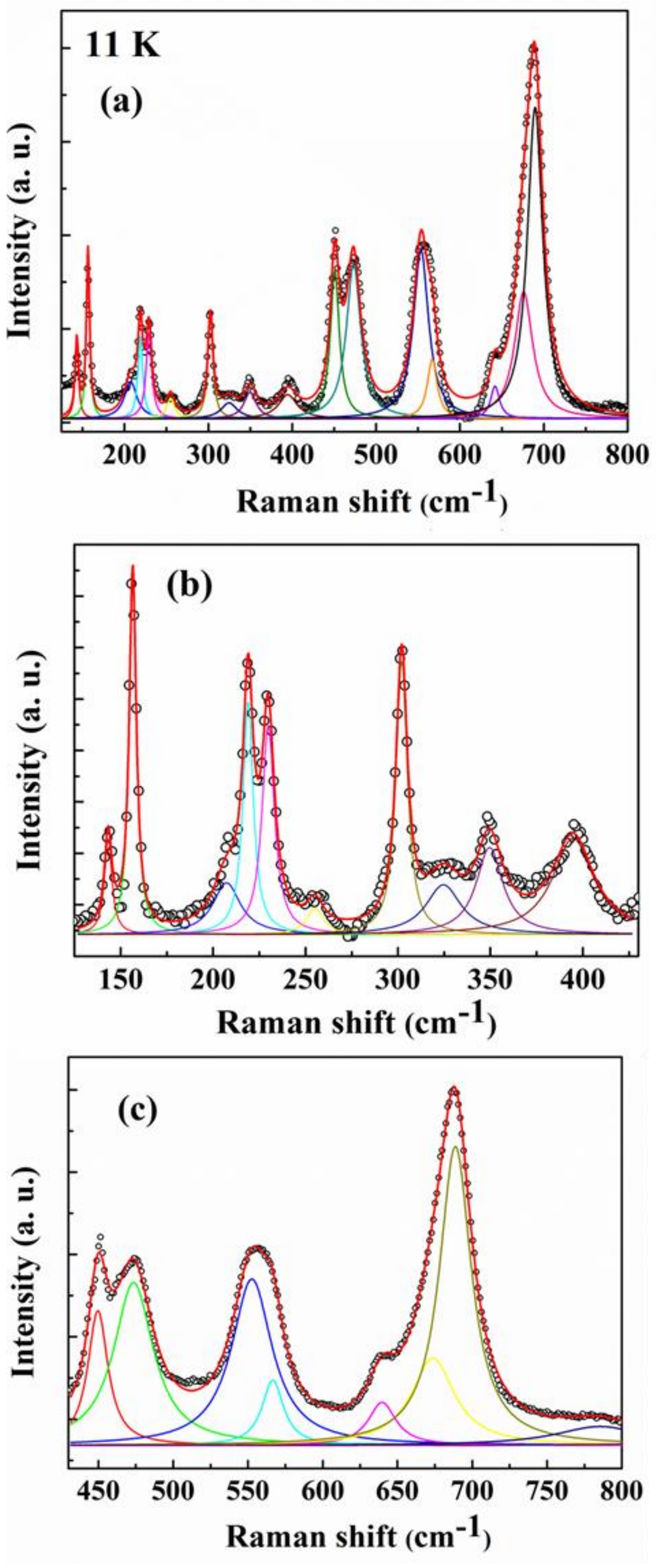

Figure 6: Fitting of Raman modes in (a) 100-800 $cm^{-1}$. (b) and (c) show the fitted modes for two different range of 100- 430 $cm^{-1}$ and 430 – 800 $cm^{-1}$ range respectively for clarity.

**Reference**

**[1]** N. R. Camara, V. T. Phuoc, I. M. Laffez, and M. Zaghriou, J Raman Spectrosc. 48, 1839 (2017).

**[2]** S. Saha, S. Chanda, A. Dutta, and T. P. Sinha, J Sol-Gel Sci Technol. **69**, 553 (2014).

**[3]** S. Chanda, S. Saha, A. Dutta, and T.P. Sinha, Mater. Res. Bull. **48**, 1688 (2013).

**[4]** K. D. Singh, R. Pandit, and R. Kumar, Solid State Sci. **85**, 70 (2018).

**[5]** M. C. Weber, J. Kreisel, P. A. Thomas, M. Newton, K. Sardar, and R. I. Walton, Phys. Rev. B **85**, 054303 (2012).